\documentclass[10pt,compsoc]{IEEEtran}
\usepackage[numbers,sort&compress]{natbib}

\usepackage[utf8]{inputenc} 
\usepackage[T1]{fontenc}    
\usepackage{hyperref}       
\usepackage{url}            
\usepackage{booktabs}       
\usepackage{amsfonts}       
\usepackage{nicefrac}       
\usepackage{microtype}      
\usepackage{lipsum}
\usepackage{threeparttable}
\usepackage{comment}
\usepackage{footnote}
\usepackage{graphicx}
\usepackage{xcolor}
\usepackage{bm}
\usepackage{makecell}
\usepackage{caption}
\usepackage{subcaption}
\usepackage{multirow}
\usepackage{tikz}
\usepackage{pifont}
\newcommand{\cmark}{\ding{51}}%
\newcommand{\xmark}{\ding{55}}%

\newcommand{\theadb}[1]{\begin{tabular}{@{}c@{}}#1\end{tabular}}

\title{Towards a Fair Comparison and Realistic Evaluation Framework of Android Malware Detectors based on Static Analysis and Machine Learning}

\begin{document}

	\author{Borja~Molina-Coronado,
		Usue~Mori,
		Alexander~Mendiburu,
		and~Jose~Miguel-Alonso
		
		\IEEEcompsocitemizethanks{\IEEEcompsocthanksitem{Borja~Molina~Coronado, Alexander~Mendiburu and Jose~Miguel-Alonso are with the Dept. of Computer Architecture and Technology, University of the Basque Country UPV/EHU, Donostia, Spain.\protect\\
		E-mail:\{borja.molina, alexander.mendiburu, j.miguel\}@ehu.es}
		\IEEEcompsocthanksitem{Usue~Mori is with the Dept. of Computer Science and Artificial Intelligence, University of the Basque Country UPV/EHU, Donostia, Spain.\protect\\
		E-mail: usue.mori@ehu.es}}%
		
	}

\IEEEtitleabstractindextext{%
	\begin{abstract}z
		As in other cybersecurity areas, machine learning (ML) techniques have emerged as a promising solution to detect Android malware. In this sense, many proposals employing a variety of algorithms and feature sets have been presented to date, often reporting impresive detection performances. However, the lack of reproducibility and the absence of a standard evaluation framework make these proposals difficult to compare. In this paper, we perform an analysis of 10 influential research works on Android malware detection using a common evaluation framework. We have identified five factors that, if not taken into account when creating datasets and designing detectors, significantly affect the trained ML models and their performances. In particular, we analyze the effect of (1) the presence of duplicated samples, (2) label (goodware/greyware/malware) attribution, (3) class imbalance, (4) the presence of apps that use evasion techniques and, (5) the evolution of apps. Based on this extensive experimentation, we conclude that the studied ML-based detectors have been evaluated optimistically, which justifies the good published results. Our findings also highlight that it is imperative to generate realistic experimental scenarios, taking into account the aforementioned factors, to foster the rise of better ML-based Android malware detection solutions.  	
	\end{abstract}

	
}
	
	\maketitle
	\IEEEdisplaynontitleabstractindextext
	
\IEEEraisesectionheading{\section{Introduction}}
	
	Over the past decade we have witnessed impressive advances in mobile devices. Along with the hardware, operating systems (OS) designed for the mobile market have experienced pairwise functionality improvements. With a market share near 72\% as of the last quarter of 2021 \cite{androidshare}, the Android platform is the leading mobile OS. It has an open source nature and is available for multiple processor architectures. These facts, along with the availability of a well documented development framework that enables a rich set of services (voice and image recognition, contactless payments, etc.), have contributed to the adoption of Android beyond smartphones \cite{alaa2017review,macario2009vehicle,do2017data}. 
	
	Coupled with this growing popularity, the increasing attention paid by malware writers to this OS has highlighted the risks to which users are exposed \cite{xu2016toward}. Data stored on mobile devices is of vital importance, sensitive for users, and has become a valuable target for attackers. According to Kaspersky, adware and banking malware targeting these devices were two of the main security threats in 2020, even being detected in trusted app markets such as Google Play \cite{Kaspersky}.
	
	Aware of these problems, researchers have seen machine learning (ML) techniques as a promising solution for the implementation of Android malware detectors \cite{tan2015securing}. ML methods leverage on app data to identify signals that are useful for detecting malware. To this end, malware detection can follow one of these strategies (or both in combination): (1) \textit{anomaly-based} detection focuses on building profiles from goodware so that deviations from those profiles are flagged as dangerous; (2) \textit{misuse-based} detection, instead, focuses on learning the characteristics of both malware and goodware in order to identify their presence in new apps \cite{molinacoronado2020survey}. Most of the research works on Android malware detection belong to the second group. More specifically, relying on supervised ML algorithms for their detection mechanisms \cite{faruki2014android,tam2017evolution}. 
	
	To perform the detection of malware using ML algorithms (either for misuse or anomaly-based systems), apps need to be preprocessed in order to extract the set of features that best describe their behavior. This task is performed using dynamic or static software analysis techniques. In \emph{dynamic analysis}, the behavior of an app is monitored in a controlled environment (sandbox), where user and system interactions are simulated. \emph{Static analysis} is based on the inspection of the files contained in the application package (APK) without needing to run the code. These techniques have their own advantages and drawbacks. On the one hand, through dynamic analysis, it is possible to access the code that is loaded and executed in runtime. However, the success of this analysis, in terms of coverage of code, greatly depends on the simulation mechanism and the absence of sandbox evasion instruments in the apps \cite{afianian2019malware}. On the other hand, static analysis is able to evaluate all the information present in the APKs, but its success lies in the absence of evasion techniques such as obfuscation or dynamic loading of code \cite{moser2007limits}. Both analysis approaches are complementary and can be combined for Android malware detection \cite{spreitzenbarth2015mobile,arshad2018samadroid}. 
	
	One of the main difficulties faced by researchers when proposing, developing and testing Android malware detectors is the absence of a common and realistic evaluation framework. This framework should include appropriate and labeled datasets, essential for training and/or testing ML algorithms. In particular, and as we illustrate in this paper, most publicly available datasets are obsolete, only contain malware, repeated samples, or comprise an insufficient number of samples. In view of this, authors opt for building ad-hoc, custom datasets by downloading app samples from different sources and labeling them using tools such as VirusTotal\footnote{\url{https://www.virustotal.com}}. This process is not only expensive, but also complicates the reproducibility and comparison of ML-based malware detection proposals.  
	
	The issue of reproducibility is aggravated by the unavailability of the code that implements the proposed methods, or by the omission in their respective publications of important details that allow their implementation. This is specially manifested for methods using ML algorithms that require tuning a large number of parameters in order to perform properly \cite{pendlebury2019tesseract}, as this information is not often provided. The same is true for the evaluation procedures. In many cases, they are not clearly described or are designed assuming very optimistic scenarios \cite{allix2016empirical, pendlebury2019tesseract}.
	
	The main objective of this study is to perform a fair comparison of Android malware detection proposals already published in the literature, shedding light about their actual effectiveness. Given the vast amount of proposals presented over the years, as well as the absence of common and realistic evaluation criteria, performing a fair comparison of methods is not a straightforward task. We have chosen 10 popular detectors based on static analysis that use different features and ML methods, and compared them under a common evaluation framework. In many cases, a re-implementation of the algorithms used in the detectors has been required due to the lack of the original authors' implementations. The result of this extensive implementation and experimental work is, to the best of our knowledge, the most comprehensive comparative study on Android malware detection methods presented to date. The scientific contributions of this work are summarized as follows: 
	
	\begin{enumerate}
		\item  We present a number of factors that negatively affect the accuracy of Android malware detectors. In particular, we consider five conditions that are present in real life but are often ignored when proposing malware detectors: (1) datasets contain a large amount of apps that are almost identical to others; (2) there is not always agreement on what is goodware and what is malware, and some apps lack enough consensus to be considered malicious or benign;
		(3) there is more goodware than malware; (4) malware authors may resort to evasion attempts using obfuscation techniques; and, (5)  malware and goodware evolve over time. Then, we argue that it is imperative to consider all these factors when designing and evaluating Android malware detectors in order to provide realistic performance values. 
		
		\item We analyze the performance of state-of-the-art ML-based Android malware detection approaches when the above factors are taken into account. For this purpose, we selected 10 highly influential detectors that make use of static analysis techniques. We show that the outstanding performances provided by the authors of these approaches are unrealistically optimistic due to design and evaluation flaws.
		
		\item We highlight the lack of reproducibility of published work in this area. In this sense, we make the code and datasets used in this comparative work publicly available.
		
		\item We discuss future research lines in Android malware detection aiming to solve the identified design and evaluation flaws.
		
	\end{enumerate}	
	
	The structure of this paper is organized as follows. Section~\ref{relwork} reviews the most important works related to the topic of this paper. Section~\ref{static_analysis} includes basic concepts about Android apps, static analysis and the type of data that can be obtained using this technique. The state-of-the-art detectors based on supervised ML classifiers that are considered in this work are presented in Section~\ref{classifiers}. Section~\ref{datasets} describes how datasets for malware detection are typically built, their limitations (factors that should be considered when constructing them but are often ignored) and the reproducibility problems that the use of custom datasets entails. In Section~\ref{experimentalsetup}, we present our experimental setup. Section~\ref{comparative} discusses the limitations of the selected Android malware detectors based on the analysis of the factors identified in Section~\ref{datasets}. Section~\ref{realistic} discusses the characteristics that a realistic evaluation framework should take into account, as well as the challenges and future research lines in Android malware detection. We conclude this paper in Section~\ref{sec:concluding}. 
	
	\section{Related Work}\label{relwork}
	
	Android malware detection is a well-studied area in the information security literature. Despite this, only a few experimental studies have focused on analyzing what factors impact the performance of malware detectors, which is the main theme of this paper. Most of them focus their analysis on a small group of detectors with similar characteristics. To date, the most comprehensive study is \cite{pendlebury2019tesseract}. This article analyzed two different sources of bias in the evaluations of three detection algorithms. The first, known as spatial bias, comes from the differences between the proportion of samples from each class in the dataset. The second type, temporal bias, is related to the inclusion of future knowledge during model training. Experiments to test for spatial bias concluded that the ratio between classes is determinant in the results reported by the authors. Experiments on the impact of temporal bias demonstrated that models tend to misclassify malware as time passes, whereas the accuracy of goodware remains stable over time. The conclusions drawn from this work are similar to those obtained in two other previous studies, each of which considers only one detector in their experiments \cite{roy2015experimental,allix2016empirical}.
	
	The use of obfuscation techniques with malware apps to evade detection was studied in \cite{rastogi2014catch}. The work presented an analysis for ten Android commercial anti-virus products, testing them by applying different obfuscation techniques to malware. The results proved that all the analyzed tools worsened their effectiveness to detect malware with at least one type of obfuscation. This work served to highlight the weaknesses of these commercial solutions. However, given that the details of the detectors are not public, specific conclusions about how obfuscation may affect ML detectors for Android cannot be drawn from this study. Another work assessed the ability of a ML malware detector for Windows to identify packed\footnote{Developers of malware apps use \emph{packers} to evade detection and analysis. A packer tool compresses and encrypts together data, resources and the code of executable files. These elements are unpacked and executed at run time.} malware \cite{aghakhani2020malware}. An extensive set of experiments was performed to confirm whether packed samples were identified due to the presence of traces left by packers or by the behavior of the sample. The authors concluded that ML detectors relying on static analysis features tend to focus on signs of obfuscation. Thus, putting in question the feasibility of these approaches against Windows malware due to the amount of false positives. Nonetheless, further studies need to be conducted to evaluate if these findings also apply to Android malware detectors.
	
	The presence of duplicates in datasets when designing and evaluating detectors is partially discussed in \cite{surendran2021impact}. Preliminary experiments are carried out for two ML detectors, one based on the usage of API calls and one using permissions. The authors of that work postulate towards the existence of a relation between duplicates and the obtention of overestimated performances for models. However, additional analysis would be required to confirm this hypothesis, as the implicit reduction of the size of the dataset for extreme duplicate removal configurations (using ample similarity thresholds) may eventually originate similar results.
	
	In summary, previous works have focused on the analysis of some specific evaluation flaws affecting detectors, including: spatial bias \cite{allix2015your,roy2015experimental,pendlebury2019tesseract}, temporal bias \cite{allix2015your,pendlebury2019tesseract}, the impact of obfuscation \cite{rastogi2014catch,maiorca2015stealth} or the influence of duplicates in the data \cite{surendran2021impact}. However, these analyses were conducted over a small number of methods \cite{roy2015experimental,allix2016empirical,surendran2021impact}, in some cases, making use of similar feature sets \cite{pendlebury2019tesseract}. Some studies were not deep enough \cite{surendran2021impact} or were performed exclusively for commercial black-box detectors, so that the details about their detection mechanisms, i.e., their features or whether they are based on signatures or ML algorithms, are not disclosed and results cannot be extrapolated \cite{rastogi2014catch,maiorca2015stealth}. None of these published works has taken into consideration the bias caused due to the removal from datasets of apps that are neither clearly goodware of malware, or has analyzed the effect of using different thresholds to label an app as malware on many existing detectors. Thus, we believe that a more comprehensive analysis, that includes all the identified factors and a higher number of detectors using different features and ML methods, becomes mandatory in the Android malware detection area. Contrary to published literature, this manuscript not only provides a comparative framework that evidences the lack of realistic proposals and illustrates many extended design and evaluation biases, but also, gives recommendations and identifies future research lines towards the proposal of realistic malware detectors.
	
	\section{Background}\label{static_analysis}
	
	This section slightly introduces some background necessary to understand the rest of the manuscript. It includes descriptions about the structure of Android apps and the type of information that can be inspected through static analysis of apps to feed ML algorithms.
	
	\subsection{Android Apps}
	
	Android apps are distributed in APKs (Android Application Packages). An APK is a ZIP compressed file that contains the resources that are essential for the execution of the app on the system: the manifest data and the compiled code. 
	
	\subsubsection{Manifest Data}
	
	The manifest file of an Android app defines a set of properties and components that the app requires from the platform in order to work. This file is in XML format and it is divided into three main blocks or sections:
	
	\begin{itemize}
		
		\item The \emph{Application Components} block defines the elements of the app that interact with the OS while the app is running or when a specific action is requested to the OS by the user. These components implement background functionality (services), manage user screens and app interactions (activities), enable app interactions with other OS components or apps (broadcast receivers and intent filters) and the interfaces to share data with other processes (content providers).
		
		\item The \emph{Hardware and Software Features} block defines the OS properties and functionalities that the app requires to function. This includes software features such as backup support, user account management, input methods, etc. Hardware features include elements such as the camera, bluetooth transmission, fingerprint sensor, etc. Declaring requested features is useful, for example, to prevent an app from running on a phone that does not fulfill the required specifications.
		
		\item The \emph{Permissions} block indicates the features that are required by an app but are protected by the OS. Access to these functionalities must be explicitly granted by the user. By default, Android apps do not have permissions to perform actions that could compromise the OS, user information, or other apps. Thus, permissions are needed to access the microphone, camera, contact list, Internet connection, location, etc.
		
	\end{itemize}

	\subsubsection{Application Code}
	
	Android apps are generally developed in Java or Kotlin (Google's preferred language for Android development) and transformed into Dalvik bytecode format during the compilation process. The Dalvik bytecode runs on the Android virtual machine, which serves as a platform-independent environment. Interactions between the hardware components of mobile devices, directly managed by the operating system, and apps, managed by the virtual machine, are performed through API libraries. These APIs provide a common way to access to the hardware capabilities required by the apps. Thus, abstracting the programmer from the particularities of devices.
	
	The Dalvik bytecode of apps is located inside the APK in the classes.dex file. It includes all user-defined classes and functions, as well as constants and variable definitions. External libraries, such as the Android framework, are not part of the content of this file. 
	
	\subsection{Static Analsis of Android Apps}
	
	Static analysis is a software technique that inspects apps to extract their characteristics without the need to execute their code and monitor their behavior at runtime \cite{li2017static}. By means of static analysis, all execution paths present in the code and all the information in the files of an app can be inspected. This is done by using tools with code interpretation mechanisms that extract understandable structures describing the internal functions of apps, e.g., call graphs, data flows, statistical measures about code structure, etc. Such information is then converted into a common set of explanatory features that will later be processed by ML algorithms.
	
	To perform the static analysis of the content of APK files, tools such as AXMLprinter\footnote{\url{https://github.com/tracer0tong/axmlprinter}} can be used to make fields and components declared inside the manifest file accessible. Instead, the classes.dex file can be converted to higher level format using decompilers such as \textit{bakSmali}\footnote{\url{https://github.com/JesusFreke/smali}}. After that, different features can be obtained, including information about instructions, methods, classes, strings and the usage of API calls \cite{faruki2014android}. It is also possible to build different graph structures representing the code. These include Call Graphs, built following the call instructions (\texttt{invoke}) present in the code; and Control Flow Graphs, which are created considering also the jumps in the code caused by conditional and loop statements (\texttt{if, switch, for, while}...). In both types of graphs, a node represents a method or a block of instructions that can only be executed sequentially, i.e., a basic block; and the edges represent the execution flow between nodes \cite{li2017static}. 
	
	After performing static analysis, the data obtained from APKs is mapped into feature vectors that represent the apps in a structured way, suitable for processing by ML algorithms. Figure~\ref{fig:bin} depicts the binary mapping as applied to represent three apps by means of their strings, API calls and permissions. It uses the values 1 or 0 to denote whether a feature is present in an app's code or not. Other mappings are also possible, for example, frequency encoding accounts for the number of times a feature is present in the app code. 
	
	\tikzset{every picture/.style={line width=0.75pt}} 
	
	\begin{figure*}[tb]
		\centering
		\vspace{1em}
		\begin{minipage}{0.7\linewidth}
			\centering
			\resizebox{0.8\linewidth}{!}{%
				\begin{tikzpicture}[x=0.75pt,y=0.75pt,yscale=-1,xscale=1]
					uncomment if require: \path (0,100); 
					
					\draw   (270,60) -- (480,60) -- (480,230) -- (270,230) -- cycle ;
					\draw   (40,60) -- (250,60) -- (250,230) -- (40,230) -- cycle ;
					\draw   (500,60) -- (710,60) -- (710,230) -- (500,230) -- cycle ;
					
					\draw (275,150) node [anchor=north west][inner sep=0.75pt]  [font=\footnotesize,rotate=-270] [align=left] {\textbf{APP2}};
					\draw (330,80) node [anchor=north west][inner sep=0.75pt]  [font=\normalsize] [align=left] {\textcolor{orange}{http://m4lw4.re}};
					\draw (350,100) node [anchor=north west][inner sep=0.75pt]  [font=\normalsize] [align=left] {\textcolor{orange}{0x3d93cb}};
					\draw (330,120) node [anchor=north west][inner sep=0.75pt]  [font=\normalsize] [align=left] {openConnection()};
					\draw (360,140) node [anchor=north west][inner sep=0.75pt]  [font=\normalsize] [align=left] {exec()};
					\draw (325,160) node [anchor=north west][inner sep=0.75pt]  [font=\normalsize] [align=left] {getConnectionInfo()};
					\draw (330,180) node [anchor=north west][inner sep=0.75pt]  [font=\normalsize] [align=left] {getInstalledApps()};
					\draw (345,200) node [anchor=north west][inner sep=0.75pt]  [font=\normalsize] [align=left] {\textcolor{blue}{INTERNET}};
					
					\draw (45,150) node [anchor=north west][inner sep=0.75pt]  [font=\footnotesize,rotate=-270] [align=left] {\textbf{APP1}};
					\draw (115,80) node [anchor=north west][inner sep=0.75pt]  [font=\normalsize] [align=left] {\textcolor{orange}{SUBSCRIBE}};
					\draw (110,100) node [anchor=north west][inner sep=0.75pt]  [font=\normalsize] [align=left] {getSmsDetails()};
					\draw (115,120) node [anchor=north west][inner sep=0.75pt]  [font=\normalsize] [align=left] {getDeviceId()};
					\draw (103,140) node [anchor=north west][inner sep=0.75pt]  [font=\normalsize] [align=left] {getLine1Number()};
					\draw (115,160) node [anchor=north west][inner sep=0.75pt]  [font=\normalsize] [align=left] {getDeviceId()};
					\draw (120,180) node [anchor=north west][inner sep=0.75pt]  [font=\normalsize] [align=left] {\textcolor{blue}{READ\_SMS}};
					\draw (120,200) node [anchor=north west][inner sep=0.75pt]  [font=\normalsize] [align=left] {\textcolor{blue}{SEND\_SMS}};
					
					\draw (505,150) node [anchor=north west][inner sep=0.75pt]  [font=\footnotesize,rotate=-270] [align=left] {\textbf{APP3}};
					\draw (580,80) node [anchor=north west][inner sep=0.75pt]  [font=\normalsize] [align=left] {\textcolor{orange}{0x3d93cb}};
					\draw (570,100) node [anchor=north west][inner sep=0.75pt]  [font=\normalsize] [align=left] {\textcolor{orange}{wallpaper\_dev}};
					\draw (570,120) node [anchor=north west][inner sep=0.75pt]  [font=\normalsize] [align=left] {getDeviceId()};
					\draw (570,140) node [anchor=north west][inner sep=0.75pt]  [font=\normalsize] [align=left] {setWallpaper()};
					\draw (575,160) node [anchor=north west][inner sep=0.75pt]  [font=\normalsize] [align=left] {\textcolor{blue}{INTERNET}};
					\draw (555,180) node [anchor=north west][inner sep=0.75pt]  [font=\normalsize] [align=left] {\textcolor{blue}{SET\_WALLPAPER}};
					
				\end{tikzpicture}
			}%
			\vspace{1.em}
		\end{minipage}
		
		\begin{minipage}{\linewidth}
			\centering
			\resizebox{\linewidth}{!}{%
				\begin{tabular}{l|c|c|c|c|c|c|c|c|c|c|c|c|c|c|c|c}
					& wallpaper\_dev & http://m4lw4.re & 0x3d93cb & SUBSCRIBE & getInstalledApps & openConnection & getSmsDetails & getDeviceId & setWallpaper & getLine1Number & getConnectionInfo & exec & SET\_WALLPAPER & INTERNET & SEND\_SMS & READ\_SMS \\
					\hline 
					\textbf{{\small APP1}} & 0 & 0 & 0 & 1 & 0 & 0 & 1 & 1 & 1 & 0 & 1 & 0 & 0 & 0 & 1 & 1 \\
					\hline 
					\textbf{{\small APP2}} & 0 & 1 & 1 & 0 & 1 & 1 & 0 & 0 & 0 & 0 & 0 & 1 & 0 & 1 & 0 & 0 \\
					\hline
					\textbf{{\small APP3}} & 1 & 0 & 1 & 0 & 0 & 0 & 0 & 0 & 1 & 1 & 0 & 0 & 1 & 1 & 0 & 0 \\
					\hline
				\end{tabular}
			}%
			
			\caption{Binary encoding of the strings (in orange), API calls (in black) and permissions (in blue) for three apps}
			\label{fig:bin}
		\end{minipage}
	\end{figure*}
	
	\section{Android Malware Detectors Based on Supervised Classifiers}\label{classifiers}
	
	Supervised classification is a popular ML task in which the objective is to learn a mapping or \emph{classifier} $\hat{H}: \mathcal{X} \to \mathcal{C}$, where $\mathcal{X}$ is a space of features that describes the samples (the input), and $\mathcal{C}$ is the space of class labels (the output). To do so, the ML algorithm is fed with a set of labeled samples $D=\{(\bm{x^1},c^1),...,(\bm{x^n},c^n)\}$ called the training set. In the context of Android malware detection using ML classifiers, the training set consists of a set of labeled apps. Each app or sample is described as a vector of $t$ features $\bm{x^i} = (x_1,x_2,...,x_t)$ which, in the context of this work, are computed through the static analysis of its APK file. The binary class label $c^i$ takes a value of $0$ for goodware apps and a value of $1$ for malware apps. Once the classifier is trained, given the feature vector of a new app $\bm{x^k}$, it will return its predicted class label $\hat{c}^k$ ($0$ for goodware and $1$ for malware).

	\begin{table*}[bt]
		
		\small
		\centering
		\caption{Android malware detection methods included in this analysis. They are the most relevant works according to the literature}\label{tabla:methods}
		\resizebox{\linewidth}{!}{%
			\begin{threeparttable}

				\begin{tabular}{c|c|c|c|c|c|c|c}
					
					
					\textbf{Method} & \textbf{\theadb{Pub.\\ Year}} & \textbf{\theadb{Dataset\\ (\#samples)}} & \textbf{APK Features} & \textbf{Encoding} & \textbf{\theadb{Dim.\\ Reduction}} & \textbf{\theadb{ML\\Algorithm}} & \textbf{\theadb{Reported\\Performance}} \\ \hline
					
					AndroDialisys \cite{feizollah2017androdialysis} & 2017 &  
					\theadb{Google Play\tnote{1} ($1\,846$) \\ Drebin ($5\,560$) } 
					& \theadb{Permissions \\ Intent Filters} & Binary & - & \theadb{Bayesian\\ Network} & \theadb{TPR: $0.955$\\FPR: $0.044$} \\ \hline
					
					BasicBlocks \cite{allix2016empirical} & 2016 &  
					\theadb{Google Play\tnote{1} ($52\,000$) \\ Android Genome ($1\,260$)} 
					& \theadb{Basic Blocks} & Binary & Mutual Info & \theadb{Random\\ Forest} & \theadb{TPR: $0.91$\\Precision: $0.94$} \\ \hline
					
					
					Drebin \cite{arp2014drebin} & 2014 &
					\theadb{Benign\tnote{1} ($123\,453$) \\ Drebin ($5\,560$)} 
					& \theadb{Permissions \\ App. Components \\ Hw. Components \\ Intents \\ Strings \\ API calls} & Binary & - & SVM & \theadb{TPR: $0.939$\\FPR: $0.01$} \\ \hline
					
					DroidDet \cite{zhu2018droiddet} & 2018 &
					\theadb{Google Play\tnote{1} ($1\,065$) \\ VirusShare\tnote{1} ($1\,065$)} 
					& \theadb{Permissions \\ Intents \\ API calls } & Binary & tf-idf rank & \theadb{Rotation\\ Forest} & \theadb{TPR: $0.884$\\Precision: $0.886$} \\ \hline
					
					DroidDetector \cite{yuan2016droiddetector} & 2016 &
					\theadb{Google Play\tnote{1} ($20\,000$) \\ Android Genome ($1\,260$) \\ Contagio\tnote{1} ($500$)} 
					& \theadb{Permissions \\ API calls } & Binary & Mutual Info\tnote{2} & DBN & \theadb{TPR: $0.981$\\FPR: $0.201$} \\ \hline
					
					
					HMMDetector \cite{canfora2016hmm} & 2016 &
					\theadb{Google Play\tnote{1} ($5\,560$)\\ Drebin ($5\,560$)} 
					& \theadb{Opcodes} & Sequence & HMM & \theadb{Random\\ Forest} & \theadb{TPR: $0.968$\\Precision: $0.96$} \\ \hline
					
					ICCDetector \cite{xu2016iccdetector} & 2016 &
					\theadb{Google Play\tnote{1} ($12\,026$)\\ Drebin\tnote{1} ($5\,264$) } 
					& \theadb{Intents \\ App. Components } & \theadb{Binary and \\ frequencies} & Mutual Info & SVM & \theadb{TPR: $0.931$\\FPR: $0.006$} \\ \hline
					
					MaMaDroid \cite{onwuzurike2019mamadroid} & 2019 &
					\theadb{Google Play\tnote{1} ($8\,447$) \\ Drebin ($5\,560$) \\ VirusShare\tnote{1} ($29\,933$)} 
					& \theadb{API call graph} & Frequencies & - & \theadb{Random\\ Forest} & \theadb{TPR: $0.97$\\Precision: $0.95$} \\ \hline
					
					MultimodalDL \cite{kim2018multimodal} & 2018 & 
					\theadb{Google Play\tnote{1} ($20\,000$)\\ Android Genome ($1\,260$)\\ VirusShare\tnote{1} ($20\,000$)} 
					& \theadb{App. Components\\Intents\\Hw. Components\\Permissions\\Opcodes\\Strings\\API calls} & Binary & - & \theadb{DL} & \theadb{TPR: $0.99$\\Precision: $0.98$} \\ \hline
					
					PermPair \cite{arora2019permpair} & 2019 & 
					\theadb{Benign\tnote{1} ($6\,993$) \\ Android Genome ($1\,264$) \\ Drebin\tnote{1} ($2\,764$)\\ Contagio\tnote{1} ($250$) \\ Koodous\tnote{1} ($2\,975$) \\ PwnZen\tnote{1} ($300$)} 
					& \theadb{Permissions} & Frequencies & \theadb{Sequential\\removal} & \theadb{Nearest\\ Neighbors} & \theadb{TPR: $0.951$\\FPR: $0.042$} \\ \hline

				\end{tabular}
				
				\begin{tablenotes}
					\item[1] Unspecified set or labeling criteria from this source.
					\item[2] In the original paper, authors use a filtered set of API calls without stating the criterion used to extract this set. As this information is unavailable, we decided to use Mutual Information to select the most important API calls based on the training data.
					
				\end{tablenotes}
				
			\end{threeparttable}
		}%
	\end{table*}
	
	As mentioned, for the purpose of this work we selected 10 malware detection methods for Android. Eight of them are selected because they have been published in in top-tier journals and represent the most important papers in this area of study, in terms of relevance, according to IEEExplore, Scopus and Google Scholar. We have also added two additional proposals, which have been considered in other experimental comparative works, namely the Drebin \cite{arp2014drebin} and BasicBlocks \cite{allix2016empirical}. All of these are \emph{misuse-}based detectors, using different supervised ML classification algorithms and with different sets of features extracted from the apps. In addition, given the large number of features they obtain from APKs, some approaches apply dimensionality reduction algorithms \cite{molinacoronado2020survey}. 
	
	Table~\ref{tabla:methods} outlines the key aspects of these detectors. As can be seen, they were published between 2014 and 2019. Due to the lack of standard datasets, an aspect which is studied in depth in the next section, they all used custom collections of apps to perform their experiments. In many cases, for these proposals, sample selection, class ratios or labeling criteria not only vary, but are not properly documented also. This makes it difficult to reproduce the experiments, compare proposals or measure their contribution level. Furthermore, none of their authors, excepting those of HMMDetector and MaMaDroid, have made their code publicly available, hindering the possibility to re-run the detectors, and complicating their use for comparison purposes. Because of these, a direct comparison of the performances reported in the corresponding articles does not provide useful information.
	
	Given the unavailability of the code of most of the detectors, we have had to re-implement them in order to perform the experimental analysis contained in this work. We tried to reimplement every proposal in the best possible way, so that we can offer fair and accurate comparisons. However, this was a complex task, mainly due to the lack of details concerning crucial aspects such as parameter values, the feature extraction and training processes of the classifier, etc. The most problematic approaches in this sense have been MultimodalDL and PermPair. Unfortunately, we were unable to implement MultiModalDL \cite{kim2018multimodal}, due to its complexity and the omission of information regarding feature computations (number of centroids, thresholds for similarity computation). As for PermPair \cite{arora2019permpair}, we implemented the detector as indicated in the original publication. However, the results obtained by us are far from those originally reported by the authors. For these reasons, we omit these works from subsequent analysis. 
	
	We have implemented the remaining eight malware detectors listed in Table~\ref{tabla:methods} in Python language\footnote{In the case of DroidDetector \cite{yuan2016droiddetector}, we were able to implement the detector assuming that feature selection was used to identify the set of most relevant API calls used by the algorithm.}. Note that we have carried out the necessary static analysis tasks on APKs to obtain the feature sets specific to each detector, as explained in Section~\ref{static_analysis}. For this purpose, we used the \emph{Androguard} framework \cite{androguard}, a widely-used static analysis and reverse engineering tool for Android APK files. For dimensionality reduction, ML algorithms and the assessment of the performance of detectors, we employed well-established libraries such as \emph{scikit-learn} \cite{scikit-learn} and \emph{numpy} \cite{harris2020array}. Our goal is not only to perform a comparative analysis between Android malware detectors, but also to contribute to the reproducibility and progress in the area by releasing our programs and the instructions for their use in our GitLab repository\footnote{To be published after acceptance}.

\section{Datasets for Android Malware Detection. Drawbacks and Reproducibility Issues}\label{datasets}

When building a supervised classifier, the dataset and the procedures used for training and evaluation are key factors in order to develop robust and well-performing models \cite{hastie2009elements}. Consequently, for ML-based malware detection, the collection of apps and the process for their generation are particularly important, both in terms of applicability to real-world scenarios and reproducibility. In this section, we identify and discuss a set of factors that must be considered when creating datasets. We then describe the datasets most commonly used by the Android malware detection research community, pointing out their main drawbacks from the point of view of the identified factors.

\subsection{Factors Under Analysis}\label{factors}

We have identified five factors that have a major impact on the performance of ML based malware detectors for Android, and that should be considered when creating the datasets used to train and evaluate them. It should be noted that, in our opinion, these factors are very important, but this does not imply that they are the only influential aspects in the design and evaluation of malware detectors based on supervised classification.

\subsubsection{REDUNDANCY}

The purpose of Android malware detectors is to identify every type of malware regardless of their level of incidence. This includes a wide range of samples pertaining to different malware families and subfamilies. Typically, malware samples within a family or subfamily exhibit analogous code and data structures to perform the same malicious activities \cite{wei2017deep}. This characteristic of malware results in datasets with groups of samples that are very similar from the point of view of detectors. This means that, within these groups, samples tend to be represented using identical feature sets, resulting in redundancies in the data fed to ML models. On the one hand, redundant samples in the training set cause bias in ML algorithms because decisions towards groups with many representatives have a great impact in the accuracy of the model \cite{hastie2009elements}. This makes models to ignore non-redundant samples, since predictions for these instances yield little improvement in accuracy. Resulting, thus, on detectors with a limited ability to identify uncommon (less represented) forms of malware. On the other hand, the presence of duplicates in the evaluation set leads to inflated or poor performances, depending on whether or not these large groups of similar apps are correctly classified. Even if the presence of duplicates or very similar apps has an important effect on the results obtained by a detector, to the best of our knowledge, only preliminary work has considered it as an issue \cite{surendran2021impact}. As a matter of fact, none of the detectors considered in our comparative analysis were assessed taking into account redundancies in the data as an major source of bias.

\subsubsection{LABELING-GREYWARE}

One of the main problems when building datasets for Android malware detectors based on supervised classification lies in the need to have labeled samples. The true nature of an app is not known in many cases. This implies that apps need to be analyzed in order to assign them a label concerning their maliciousness. Ideally, labeling should be carried out manually, by experts, to guarantee the highest number of error-free labels. However, using human annotators to perform this task is costly, both in terms of time and resources \cite{li2017android} and it is not error-free either. Accordingly, researchers usually rely upon automatic procedures to label their customized datasets.

In the simplest labeling approach, apps are labeled according to the source they were obtained from \cite{arora2019permpair}, e.g., those downloaded from trusted repositories (such as Google Play) are goodware, whereas those from repositories such as VirusShare are labeled as malware. This procedure relies on the analysis, either automatic or manual, performed by the managers of these repositories. In other cases, apps are scanned using a collection of antivirus programs and labeled depending on the number of positive (malware) alerts \cite{souri2018state}. In this context, VirusTotal is a tool that allows users to upload files, including APKs, and scan them using a collection of antivirus engines. VirusTotal results, based on the number of positive alerts raised by a file (we refer to this number as VTD, from VirusTotal Detections), are widely used as the criterion to label samples. To this end, a common procedure makes use of thresholds to establish the level of consensus required to label an APK as malware or goodware \cite{salem2019don}.

Leveraging on the VTD leads authors to decide what threshold is adequate for malware and goodware, so it is common to find disparities in the literature. For example, \cite{roy2015experimental} set a threshold of VTD$\geq$10 to flag an app as malware, while in \cite{pendlebury2019tesseract}, authors set this threshold at a much lower value of 4. In both cases, the condition for labeling an app as goodware was set to VTD$=$0. The choice of thresholds not only means that the malware or goodware definitions vary among articles, but also influences the amount of apps that fall between these categories. These apps, which we refer to as greyware, lack enough consensus by antivirus programs to be considered as goodware or malware with guarantees. Because of that, many researchers discard these apps when training their detectors. However, greyware is an important part of the Android ecosystem \cite{AndroidSec} and will appear whenever the detector is deployed in a real environment. Thus, discarding them at training time will hinder the effectiveness of a detector once deployed. Indeed, even if common thresholds were applied to label the data, the VTD value provided by VirusTotal changes over time \cite{zhu2020measuring}, for example, due to engine updates aimed at improving detection capability, or as engines are added or removed from the platform. Consequently, disparities may occur between models validated with exactly the same collection of apps but labeled at different times \cite{salem2019don}.

\subsubsection{IMBALANCE}

The third factor we consider is related to the ratio of malware and goodware in the dataset, especially in the data used for training. In real life, most apps are innocuous in the security aspect, with the actual proportion of malware being about 10\% of the total number of Android apps \cite{pendlebury2019tesseract}, but with this ratio being highly dependent on the particular market from which apps are downloaded \cite{googlereport,lindorfer2014andradar}. As can be seen in Table~\ref{tabla:methods}, researchers have trained their detectors following their own criterion and assuming different class proportions. However, the choice of the class ratio is important when training ML models, since classical ML classifiers tend to be biased towards the majority class in highly unbalanced scenarios. Therefore, ignoring class imbalance can lead to a false perception about the true performance of detectors under real working conditions \cite{pendlebury2019tesseract}. 

\subsubsection{EVASION}

Some malware authors are aware that their apps will be examined by malware detectors, so they try to bypass detection by using different evasion techniques. As technology evolves, different threats may appear and authors should be aware that their systems will be the target of attacks. Thus, a proper evaluation of the robustness of malware detectors against such threats should be taken into consideration. Unfortunately, none of the proposals included in this comparative work have considered the effects of evasion techniques, such as obfuscation, in their evaluations. 

\subsubsection{EVOLUTION}

The last factor we highlight is related to the evolution of malware and goodware over time. This implies that the behavior of apps rarely remains static for a long time. Thus, the characteristics of newer or mutated apps may differ from those obtained from apps observed earlier, during the training of a detector. Despite that, most authors design and/or evaluate their ML detectors on the assumption that future malware and goodware will remain similar to that used at design time \cite{roy2015experimental,pendlebury2019tesseract}.

\subsection{Available Android Datasets and their Drawbacks}\label{currentdatasets}

Android datasets that are used for building supervised classifiers consist of collections of apps and their associated labels, which indicate whether an app is malware or goodware. We have searched in the literature for datasets of Android APKs designed for research on misuse-based Android malware detectors, and have found five popular ones. Their characteristics have been summarized in Table~\ref{tabla:datasets}, taking into account all the factors identified in the previous section. 

\begin{table*}[htb]
	\centering	
	\caption{Characteristics of popular Android Malware Datasets. The ``?'' symbol means that no information about this characteristic is reported }\label{tabla:datasets}
		
		\begin{threeparttable}
			
			\begin{tabular}{c|c|c|c|c|c|c}
				
				
				\textbf{Dataset} & \textbf{Time period} & \textbf{\theadb{Labeling method \\ (Type of labels)}} & \textbf{\theadb{\#samples}} & \textbf{\theadb{Obfuscated\\samples}} & \textbf{\theadb{Redundant\\samples}} & \textbf{\theadb{Timestamps}} \\ \hline
				
				Android Genome \cite{zhou2012dissecting} & 2010-2011 & Manual (Binary) & \theadb{$1\,260$ malware} & ? & ? & \xmark \\ \hline
				
				Drebin \cite{arp2014drebin} & 2010-2012 & VirusTotal (Binary) & \theadb{$5\,560$ malware} & ? & ? & \xmark \\ \hline
				
				AMD \cite{wei2017deep} & 2010-2016 & Hybrid (Binary) & \theadb{$24\,553$ malware} & ? & ? & \xmark \\ \hline
				
				CICAndMal2017 \cite{lashkari2018toward} & 2014-2017 & VirusTotal (Binary) & \theadb{$426$ malware\\$5\,065$ goodware} & ? & ? & \xmark \\ \hline
				
				AndroZoo \cite{allix2016androzoo}\tnote{1} & 2011- & VirusTotal (VTD) & \theadb{$13\,045\,285$ mixed} & ? & ? & \cmark \\ \hline
				
			\end{tabular}
			
			\begin{tablenotes}
				
				\footnotesize
				\item[1] This dataset is continuously growing
				
			\end{tablenotes}
			
		\end{threeparttable}
		
	
\end{table*}

In two of the reported datasets (Drebin and CICAndMal2017), the labels of the samples were obtained by setting some threshold over the VTD. For example, in Drebin, an app is tagged as malware when its VTD$\geq$2, for a subset of eight antivirus engines from VirusTotal that are selected as reliable by the authors. Only the Android Genome dataset was built based on manual labeling. A combination of both labeling approaches was used in the AMD collection: automatic labeling was first carried out using VirusTotal to filter and cluster apps into malware families, and then a small subset from each family was manually verified. Finally, note that AndroZoo does not provide labels, supplying VTD values instead --so it is up to the user how to use this information for labeling.

To properly train detectors based on ML classifiers, both malware and goodware samples are needed. Ideally, greyware should also be included. Nonetheless, Drebin, Android Genome and AMD comprise exclusively malware samples and only AndroZoo allows samples to be labeled as greyware. Another drawback of these datasets is related to obfuscated malware. In this sense, the authors neither identify, or explicitly include, obfuscated versions of malware, which makes it very difficult to analyze the possible effects of evasion attempts on the performance of detectors. Something similar occurs for redundant samples, as none of the papers describing the datasets provide information about their presence.

Finally, as can be seen in Table~\ref{tabla:datasets}, most of these datasets were created more than five years ago and may contain old-fashioned malware. The most recent dataset is CICAndMal2017, which comprises a small set of apps released between 2014 and 2017. This small number of instances may not be enough for training and evaluating anti-malware methods based on some ML algorithms \cite{raudys1991small}. In addition, in most cases the samples included in these datasets cover only a small period of time or their release date is not available. These facts complicate the elaboration of proper analysis about the evolution of the characteristics of apps. 

The limitations of these datasets have led researchers to build custom datasets, in a similar way as AndroZoo was created, i.e., by combining APKs downloaded from various sources or repositories. AndroZoo constitutes the most important public source of Android apps for researchers \cite{allix2016androzoo}. Sources of AndroZoo include app marketplaces (such as Google Play), malware datasets, torrents and different malware repositories\footnote{For example: ContagioDump (\url{https://contagiodump.blogspot.com/}), Koodous (\url{https://koodous.com}), VirusShare (\url{https://virusshare.com/}).}. Since 2011, these sources are continuously tracked to keep the collection up to date. Additionally to the APKs, AndroZoo provides a file with information about the apps contained in the dataset. The contents of this file, which is updated daily, can be used to filter the samples to be downloaded according to different criteria, such as the market from which an APK was obtained, the SHA and the date of the APK, and the VTD value.

From the above discussion, it is clear that the decisions made by the authors during the construction of custom datasets make them unique, and that particularities of these datasets influence the performance of detectors. Moreover, performance is not only dependent on the data used, but also on the design of the experiments using these data. In our opinion, and as mentioned in the previous section, the use of different datasets and the consideration of different guidelines in the design of the experiments, makes the comparison of the metrics reported for these proposals meaningless, entails an obstacle to the reproducibility and seriously impairs progress on this relevant topic.

\section{Experimental Setup}\label{experimentalsetup}

In this section we present the experimental setup devised for this work. We describe the datasets used, the process to build and evaluate the Android malware detectors and the evaluation metrics considered to measure their performance.

\subsection{Master Dataset}\label{dataset}

Our starting point is a ``master'' dataset that allows us to derive datasets useful to evaluate on the selected detectors every factor presented in Section~\ref{factors}, i.e., REDUNDANCY, LABELING-GREYWARE, IMBALANCE, EVASION and EVOLUTION.

To build our master dataset we selected apps from AndroZoo, taking into account their origin, the reported VTD and the date. We use apps from Google Play, AppChina and VirusShare. The date of the apps was used to select 100 monthly samples of each class (malware, goodware and greyware) during the period starting from January 2012 to December 2019. As we mentioned earlier, there is a lack of ground truth in the area, and automated labeling methods based on VirusTotal have become commonplace. Given the lack of standard criteria in the literature to interpret the results obtained from VirusTotal \cite{salem2019don}, in this paper we resort to a simplified but operational automated labeling method: the use of VTD values and the application of a set of pre-defined thresholds. With this method we have a practical definition of what is goodware, what is greyware and what is malware that allows us to fairly compare detectors and to show different sources of experimental bias. In particular, we used the median of the thresholds used in two previous works \cite{roy2015experimental,pendlebury2019tesseract} to label the malware, i.e., a VTD$\geq$7. Goodware samples were considered as those with VTD=0, whereas samples with a 1$\leq$VTD$\leq$6 rating were labeled as greyware. Moreover, we restricted our analysis to samples discovered at most in 2019 to ensure that malware signatures for samples in our dataset are well stabilized. Such criteria are in concordance with what is recommended in \cite{zhu2020measuring} to obtain reliable labels from VirusTotal.

In total, our master dataset consists of 28,800 samples. The complete list of APKs and the instructions to download this dataset, along with our code, are available in our GitLab repository. In contrast to the datasets introduced in Section~\ref{currentdatasets}, this dataset is more comprehensive as: (1) it considers greyware; (2) it provides a sufficiently large number of goodware, malware and greyware samples to properly train and test ML detectors; (3) it considers a larger time period and the organization of samples by months allows us to split our dataset according to the age of the apps. Additionally, compared to the datasets used for training the detectors included in this analysis (see Table \ref{tabla:methods}), the procedure followed to build our master dataset and its derived datasets is more systematic, rigorous and transparent.

\subsection{Model Training and Assessment Process}\label{evalcriteria}

For model construction and evaluation, the dataset is divided into training and test partitions. In order to obtain unbiased results, the test partition is always kept as a completely separate set and is never used for training nor for feature engineering processes (extraction, preprocessing or dimensionality reduction). The model parameters are selected using standard k-fold cross validation (with $k=5$) within the training set and following a grid search approach. In scenarios where EVOLUTION is considered because the distributions of the data are assumed to change over time, time-aware k-folds are used \cite{arlot2010survey}, i.e., all the apps of the evaluation sets are more recent than those for training. In all cases, folds are created maintaining the original ratio between the classes in the given experimental scenario. This is a common process in the ML literature for estimating the generalization error of the final model \cite{hastie2009elements}. 

\subsection{Evaluation Metrics}

For this work, we considered a set of evaluation metrics (see Table~\ref{metrics}) which are common in the ML and computer security literature \cite{molinacoronado2020survey}. Note that some of these metrics, such as the TPR, the FPR or precision, offer complementary information and should be used together to fully understand the performance of a system. These metrics make use of: true positives (TP), which indicate the number of correct positive predictions; false positives (FP), that account for the number of incorrectly predicted negative elements; and P and N, which make reference to the number of positive and negative elements in the data. Conversely, summary metrics such as A$_{mean}$, F1 score or kappa statistic, quantify the performance of a system using one measure, but lack explainability and do not provide insights regarding the causes of the good or bad performance. F1 represents the harmonic mean between precision and TPR, A$_{mean}$ is the arithmetic mean of the accuracies for the positive and negative classes, and the kappa statistic quantifies the level of correlation between the predictions made by a classifier and the actual labels in the data ($P_o$ being the accuracy of the detector, $P_c$ being the weighted sum of the predictions, $p_k$ being the actual proportion of samples of the class $k$ in the data and $\hat{p_k}$ being the proportion of samples predicted as pertaining to class $k$). Unlike the F1 score (as it does not consider True Negatives), the kappa and A$_{mean}$ metrics are especially useful under unbalanced problems.

\begin{table}[t]
	\centering
	\caption{Evaluation metrics used in this paper}
	\begin{tabular}{| c | c | c |}
		\hline
		\theadb{$$\\$ TPR = \frac{TP}{P} $\\$$} & 
		\theadb{$ FPR = \frac{FP}{N} $} &
		\theadb{$ Precision = \frac{TP}{TP+FP} $}
		
		\\ \hline
		
		\multicolumn{2}{|c|}{\theadb{$$\\$ F1 = 2*\frac{TPR*Precision}{TPR+Precision} $ \\ $$}} &
		{\theadb{$ A_{mean} = \frac{TPR+1-FPR}{2} $ }}
		\\ \hline
		\multicolumn{2}{|c}{{\theadb{$$\\$ kappa = \frac{P_o-P_c}{1-P_c} $\\$$}}} & \small{\theadb{$P_o = \frac{TP + TN}{P + N}$ \\ $P_c = \sum_{k}^{}{p_k*\hat{p}_k}$}} \\ \hline
		
	\end{tabular}
	
	\label{metrics}
\end{table}

\section{Comparative Analysis}\label{comparative}

In this section we run a complete set of experiments, designing specific scenarios to analyze the effect of the factors presented in Section~\ref{factors}. First, a basic scenario is used as the departing point for subsequent experiments. In all scenarios, the selected detectors are compared in equal conditions. 

Departing from the master dataset, two different datasets have been extracted, namely balanced and unbalanced (see Table~\ref{tab:trainconfigs}). For the balanced case, it is assumed that malware and goodware are equally probable. Thus, from the period 2012-2015, $\sim$70\% of all the goodware and $\sim$70\% of all the malware are uniformly sampled on a monthly basis and used for training, keeping the remaining 30\% of the samples for testing purposes. On the contrary, under the unbalanced configuration, malware is assumed to be less frequent than goodware and, consequently, the malware is downsampled to become $\sim$10\% of the goodware. This process is performed by randomly sampling malware on a monthly basis according to a Gaussian distribution with parameters $N(0.1,0.02)$.

\begin{table*}[tb!]
	
	\small
	\centering
	\caption{Composition of Balanced and Unbalanced datasets derived from the 2012-2015 period of the Master Dataset}\label{tab:trainconfigs}
		\begin{tabular}{l|c|c|c||c|c|c}
			
			& \multicolumn{3}{c||}{\textbf{Training}} & \multicolumn{3}{c}{\textbf{Testing}} \\
			& \textbf{\theadb{\%malware\\(\#samples)}} & \textbf{\theadb{\%goodware\\(\#samples)}} & \textbf{ratio} & \textbf{\theadb{\%malware\\(\#samples)}} & \textbf{\theadb{\%goodware\\(\#samples)}} & \textbf{ratio} \\ \hline
			
			Balanced & 70\% (3360) & 70\% (3360) & 1:1 & 30\% (1440) & 30\% (1440) & 1:1 \\ \hline
			Unbalanced & $\sim$7\% ($\sim$336) & 70\% (3360) & $\sim$1:10 & $\sim$3\% ($\sim$144) & 30\% (1440) & $\sim$1:10 \\ \hline
			
		\end{tabular}
\end{table*}

\subsection{Baseline Scenario}

Before analyzing the scenarios related to each factor, we test detectors under the most basic and unrealistic assumptions, i.e., discarding greyware, considering a balanced ratio between the classes, omitting evasion attempts and ignoring the evolution of the apps. The aim of this scenario is to mimic the (favorable) conditions that are commonly assumed in the literature and try to reproduce the results obtained in the original papers.

As shown in Table \ref{tab:baseline}, half of the detectors showed promising detection performances in this scenario, achieving detection rates (TPR) over $0.8$ and moderately low false positives of about $0.1$. The best detector under this scenario in terms of summary metrics ($0.91$ for kappa), despite being one of the first works in the area, was Drebin, with TPR and FPR values of $0.95$ and $0.04$, respectively. Another two well-performing detectors were DroidDet and MaMaDroid, with similar kappa figures of $0.84$, despite using different features and ML algorithms. The good results reported for some methods contrast with those obtained for DroidDetector, HMMDetector and AndroDialysis. Given the high number of false positives of these detectors (above 20\%), one could conclude that these models are inappropriate even under optimistic working conditions.

\begin{table}[tb]
	
	\centering
	
	\caption{Performance of detectors for the baseline scenario}
	
	\resizebox{\linewidth}{!}{%
		
		\begin{tabular}{c|c|c|c|c|c|c}
			
			\textbf{Method} & \textbf{TPR} & \textbf{FPR} & \textbf{Precision} & \textbf{F1} & \textbf{A$_{mean}$} & \textbf{Kappa} \\ \hline
			
			AndroDialysis & 0.848 & 0.291 & 0.744 & 0.792 & 0.778 & 0.556 \\ 
			
			BasicBlocks & 0.823 & 0.097 & 0.894 & 0.857 & 0.863 & 0.726 \\ 
			
			
			Drebin & 0.953 & 0.043 & 0.956 & 0.955 & 0.955 & 0.910 \\ 
			
			DroidDet & 0.936 & 0.088 & 0.913 & {0.925} & {0.924} & {0.848} \\ 
			
			DroidDetector & 0.472 & 0.335 & 0.585 & 0.523 & 0.568 & 0.137 \\ 
			
			HMMDetector & 0.824 & 0.560 & 0.595 & 0.691 & 0.631 & 0.263 \\ 
			
			ICCDetector & 0.800 & 0.093 & 0.895 & 0.845 & 0.853 & 0.706 \\ 
			
			MaMaDroid & 0.930 & 0.085 & 0.915 & {0.923} & {0.922} & {0.845} \\ \hline
			
		\end{tabular}
		
		}%

	\label{tab:baseline}
	
\end{table}

\subsection{Redundancy Scenario}

This scenario aims to study the impact of the presence of very similar samples in Android datasets. To this end, we filter out malware and goodware apps from our initial balanced dataset. This is done by computing intra-class similarities between samples and then applying the algorithm proposed in \cite{surendran2021impact}. The algorithm works by randomly selecting one sample at a time and removing the samples lying in its neighborhood, according to an $\epsilon$ radius threshold.

For our experiments, we use call frequencies of APIs as the representation of apps and the Euclidean distance to compute the degree of similarity between samples. The intuition behind the consideration of this representation lies in that API call frequencies make reference to both the set of API calls that describe the actions carried out by apps, and the prevalence of these calls in the code. Thus, we assume that when the Euclidean distance for two slightly different apps is below or equal to $\epsilon$, both apps perform identical actions, e.g., they are variations of the same app. 

\begin{figure}[t]
	\centering
	\includegraphics[width=\linewidth]{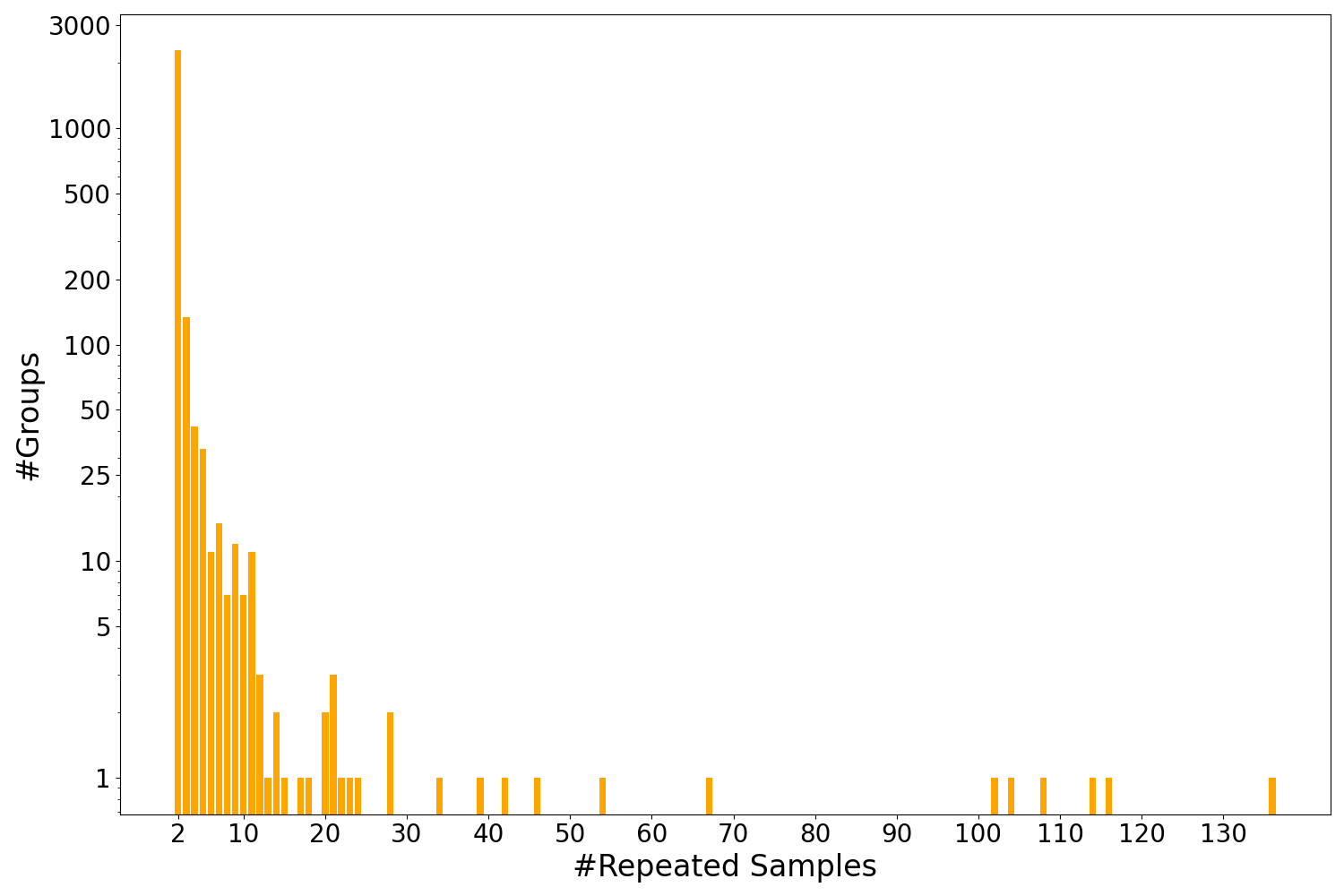}
	\caption{Number of malware groups per group size. Note the logarithmic scale in the $Y$ axis.}
	\label{fig:group_hist}
\end{figure}

We perform the filtering process using a redundancy tolerance value $\epsilon$ equal to 0, which means that we consider as duplicates two apps with \emph{exactly} the same API call frequencies. As a result of this process, a dataset without exact duplicates is obtained. A summary of the characteristics of this dataset are presented in Table~\ref{table:filtdatasets}. As can be seen, the size of the malware subset is reduced substantially (almost half of the malware was filtered out), whereas for goodware the number of removed duplicates is considerably smaller. Details about the groups found for malware during the deduplication process are shown in Figure~\ref{fig:group_hist}. It depicts the number of groups found int the malware dataset arranged by group size. As can be seen, our filtered dataset is heterogeneous in terms of unique malware behaviors. More specifically, 2285 samples only contain a duplicate, which represent 88\% of the filtered malware. Groups with a small number of duplicates represent the majority of the data, with 70\% of the original malware containing less than 10 identical samples. Large groups are also present. In this regard, about 2\% of the resultant malware groups have more than 10 duplicates.

\begin{table}
	\centering
	\caption{Composition of the full dataset used for the filtered scenario (data from 2012 to 2015)}\label{table:filtdatasets}
	\begin{tabular}{c|c|c}
		\textbf{Dataset} & \textbf{\#malware} & \textbf{\#goodware} \\ \hline
		Original & $4800$ & $4800$ \\ \hline
		Filtered & $2588$ & $4291$ \\ \hline
	\end{tabular}
\end{table}

\begin{table}[t]
	
	\centering
	
	\caption{Performance of detectors trained using the filtered dataset and evaluated with the baseline test set}
	
	\resizebox{\linewidth}{!}{%
		
		\begin{tabular}{c|c|c|c|c|c|c}
			
			\textbf{Method} & \textbf{TPR} & \textbf{FPR} & \textbf{Precision} & \textbf{F1} & \textbf{A$_{mean}$} & \textbf{Kappa} \\ \hline
			
			AndroDialysis & 0.904 & 0.365 & 0.712 & 0.796 & 0.769 & 0.538 \\ 
			
			BasicBlocks & 0.893 & 0.044 & 0.952 & 0.922 & 0.924 & 0.849 \\ 
			
			
			Drebin & 0.933 & 0.031 & 0.966 & 0.949 & 0.950 & 0.901 \\ 
			
			DroidDet & 0.933 & 0.056 & 0.942 & {0.937} & {0.938} & {0.876} \\ 
			
			DroidDetector & 0.654 & 0.403 & 0.618 & 0.635 & 0.625 & 0.250 \\ 
			
			HMMDetector & 0.796 & 0.488 & 0.619 & 0.697 & 0.653 & 0.307 \\ 
			
			ICCDetector & 0.800 & 0.053 & 0.937 & 0.863 & 0.873 & 0.747 \\ 
			
			MaMaDroid & 0.914 & 0.046 & 0.951 & {0.932} & {0.934} & {0.868} \\ \hline 
			
		\end{tabular}
		
		}%

	\label{tab:filteredtest}
	
\end{table}

\begin{table}[t]
	
	\centering
	
	\caption{Performance of detectors trained and evaluated with filtered sets} 

\resizebox{\linewidth}{!}{%
	
	\begin{tabular}{c|c|c|c|c|c|c}
		
		\textbf{Method} & \textbf{TPR} & \textbf{FPR} & \textbf{Precision} & \textbf{F1} & \textbf{A$_{mean}$} & \textbf{Kappa} \\ \hline
		
		AndroDialysis & 0.913 & 0.362 & 0.716 & 0.802 & 0.775 & 0.551 \\ 
		
		BasicBlocks & 0.880 & 0.065 & 0.930 & 0.904 & 0.907 & 0.814 \\ 
		
		
		Drebin & 0.925 & 0.056 & 0.942 & 0.933 & 0.934 & 0.868 \\ 
		
		DroidDet & 0.926 & 0.073 & 0.926 & {0.926} & {0.926} & {0.853} \\ 
		
		DroidDetector & 0.728 & 0.395 & 0.647 & 0.685 & 0.666 & 0.332 \\ 
		
		HMMDetector & 0.774 & 0.545 & 0.586 & 0.667 & 0.614 & 0.229 \\ 
		
		ICCDetector & 0.797 & 0.070 & 0.918 & 0.853 & 0.863 & 0.726 \\ 
		
		MaMaDroid & 0.900 & 0.072 & 0.925 & {0.913} & {0.914} & {0.828} \\ \hline
		
	\end{tabular}
	
	}%

\label{tab:filtered00}

\end{table}

In order to confirm the influence of the presence of very similar apps in the generalization ability of ML models, the results for detectors trained with filtered data\footnote{As in the baseline scenario, a balanced ratio between goodware and malware is used for training. This is achieved by randomly downsampling the majority class (goodware).} and evaluated with the unfiltered test set used for the baseline scenario are shown in Table \ref{tab:filteredtest}. As can be seen, almost all approaches improve their performances, with similar TPR and lower FPR values, compared to the results for detectors trained with the unfiltered training set (see Table~\ref{tab:baseline}). Among the most benefited methods in terms of performance are ICCDetector and BasicBlocks, with improvements of $5\%$ and $16\%$ in their kappa values, respectively. The results observed for this scenario support the hypothesis that duplicate samples in the training set result in biased models, since accurate predictions for these large groups at training time result in higher accuracy, and the models are less prone to pay attention to minority groups. Thus, removing duplicates from training is translated into models that are able to generalize better to different types of instances.

The previous experiment was carried out with an unfiltered test set. However, to evidence the impact that the presence of duplicates in the testing set has on the reported performances, we perform evaluations using a filtered test set. Table \ref{tab:filtered00}, depicts the performance for detectors both trained and evaluated using data without duplicates. Among all the methods, the best scoring are Drebin and DroidDet with kappa values of $0.86$ and $0.85$, respectively. However, as expected, most detectors report lower detection performances in this scenario than in the previous experiments (see Table~\ref{tab:filteredtest} and Table~\ref{tab:baseline}). With duplicates in the test set, correctly identifying bigger groups has a significant positive effect on the measured performance, whereas erroneously classifying these groups penalizes it. This effect disappears when duplicates are removed and only one representative per group is left. Therefore, the measured performance is a better indicator of generalization ability of detectors.\\

\subsection{Labeling-Greyware Scenarios}

As previously mentioned, manual, exhaustive and error-free labeling is not affordable. For this reason, the community has agreed to use tools such as VirusTotal to automatically label apps. In this sense, using thresholds on the VTD value is the most common class separation criteria. However, different interpretations and thresholds have been used, which impacts on the obtained datasets. Two interesting analyses arise here: (1) how the selection of the threshold affects performance and, (2) what is the behavior of detectors when they examine apps that were discarded during labeling. 

First, we evaluate how the selection of different criteria for labeling datasets affects detectors and thus, the reported performance. Specifically, we want to analyze whether the uncertainty and difficulty of the problem is greatly reduced when higher VTD values are used for labeling malware. To do so, models are trained similarly to the baseline scenario, i.e., using a balanced ratio between both classes with samples selected from 2012 to 2015, but varying the threshold used for labeling.

\begin{figure}[ht]
\centering
\includegraphics[width=\linewidth]{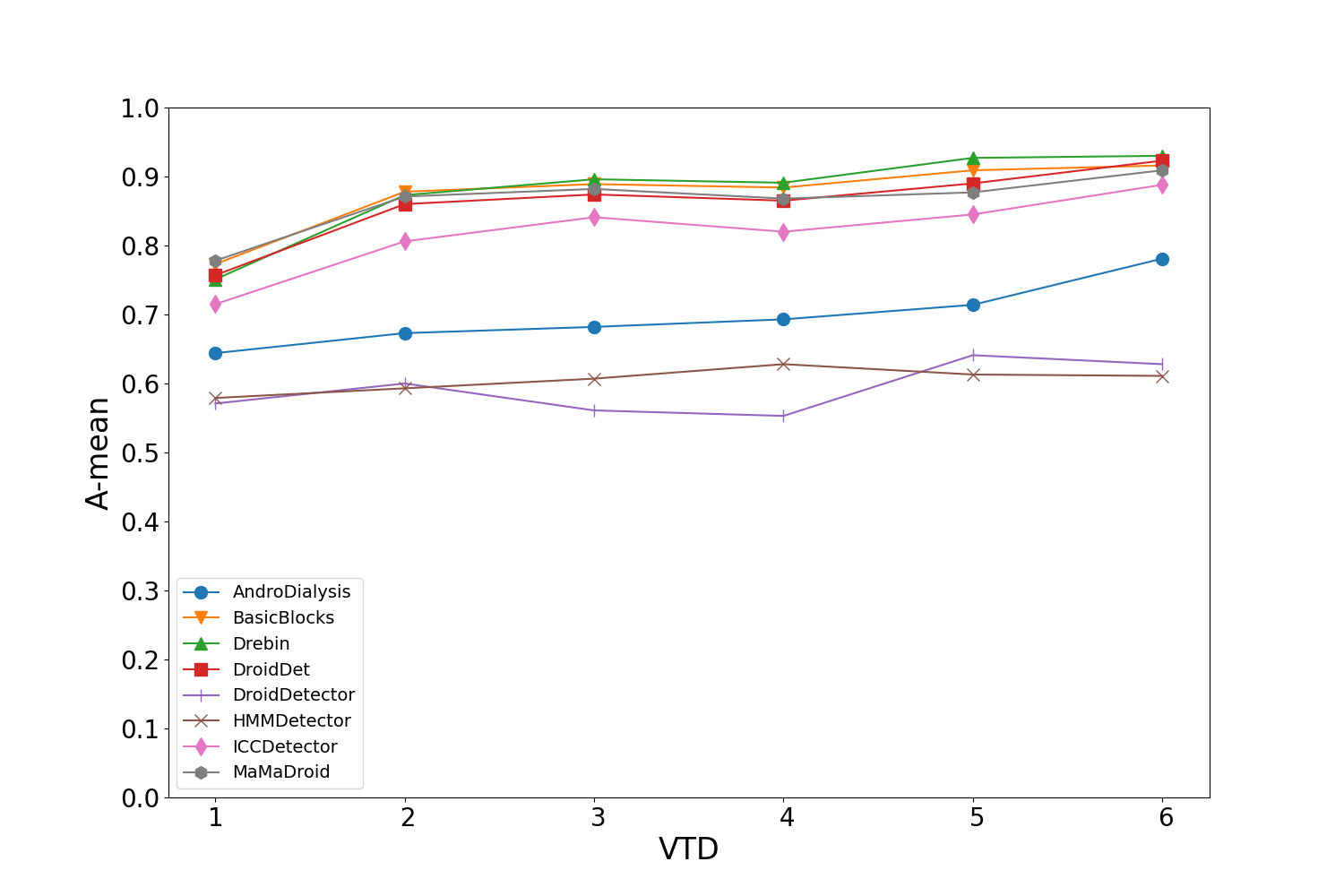}
\caption{A$_{mean}$ values for different VTD labeling thresholds}
\label{fig:amean_vtd}
\end{figure}

Results for this experiment are shown in Table~\ref{vtdvalues}. As can be seen, the TPR of models is directly correlated with the VTD used for labeling the dataset i.e., the higher the VTD, the higher the TPR of the detector. The contrary is true for the FPR, since it follows a reverse trend with respect to the VTD. This confirms that the malware detection problem becomes easier whenever the separation between both classes, in terms of the VTD, is enlarged. Figure~\ref{fig:amean_vtd} exhibits this effect for the values of A$_{mean}$ obtained for the different detectors.

The second scenario is devoted to show how detectors behave when facing greyware. Specifically, we want to analyze whether the uncertainty and difficulty of the problem increases when adding greyware and, thus, the omission of these apps results on the oversimplification of the problem and leads to unrealistic and unfair evaluations. To do so, models are trained using the balanced dataset and tested exclusively on greyware samples from the period between 2012 and 2015.

We have compiled the responses of the detectors when classifying greyware samples in Table~\ref{greyware}. G indicates how many input samples are identified as goodware, whereas M refers to how many samples are classified as malware. The total results are in the right column of the table. The other columns show the partial results for the input samples grouped by their VTD scores. As can be seen, on average 35\% of the total samples are considered goodware by detectors. As expected, the results show certain correlation between the VTD and the decisions made by the detectors: given an app, the higher its VTD is, the more likely the detectors are to classify it as malware. We can observe a high uncertainty for samples with lower VTD values, explaining why authors opt to discard greyware from their experiments. As a result, the problem solved by detectors is simplified. Thus, providing artificially boosted performance results and hiding an effect that will appear in real-working conditions.

In addition, we also analyze how the VTD value and thus, the labels, change over time. For the labeling of our dataset we used the VTD scores provided by AndroZoo. We completely reanalyzed the apps in our dataset with VirusTotal and computed the confusion matrix to represent the label swaps between the two analyses. As can be seen in Table~\ref{swaps}, most swaps occur for the greyware class. In this sense, more than half of the apps change their label from greyware to goodware (27.2\%) or malware (28.1\%). Also, 9.9\% of the goodware and 6.9\% of the malware apps fall into the category of greyware after being reanalyzed. These changes illustrate the downsides and limitations of using the VTD score for labeling. On the one hand, the need of providing the labels of samples when releasing datasets to avoid class swaps caused by the reanalysis of apps and to guarantee reproducibility. On the other hand, the need to reconsider greyware as part of the datasets, as 44\% of the apps remain in this category even after being reanalyzed.

\begin{table*}[tb]
\centering
\caption{Performance of detectors trained with data labeled using different thresholds over the VTD for the malware. TPR stands for the True Positive Ratio and FPR is the False Positive Ratio}\label{vtdvalues}
	\begin{tabular}{c|cc|cc|cc|cc|cc|cc}
		& \multicolumn{2}{c|}{\textbf{VTD>=1}} & \multicolumn{2}{c|}{\textbf{VTD>=2}} & \multicolumn{2}{c|}{\textbf{VTD>=3}} & \multicolumn{2}{c|}{\textbf{VTD>=4}} & \multicolumn{2}{c|}{\textbf{VTD>=5}} & \multicolumn{2}{c}{\textbf{VTD>=6}} \\ \hline
		\textbf{Method} & \textbf{TPR} & \textbf{FPR} & \textbf{TPR} & \textbf{FPR} & \textbf{TPR} & \textbf{FPR} & \textbf{TPR} & \textbf{FPR} & \textbf{TPR} & \textbf{FPR} & \textbf{TPR} & \textbf{FPR} \\ \hline
		AndroDyalisis & 0.547 & 0.258 & 0.622 & 0.275 & 0.549 & 0.183 & 0.692 & 0.305 & 0.798 & 0.369 & 0.861 & 0.298 \\
		BasicBlocks   & 0.731 & 0.184 & 0.869 & 0.112 & 0.885 & 0.107 & 0.904 & 0.134 & 0.909 & 0.090 & 0.916 & 0.083 \\
		Drebin        & 0.731 & 0.227 & 0.860 & 0.113 & 0.901 & 0.107 & 0.914 & 0.131 & 0.935 & 0.081 & 0.923 & 0.062 \\
		DroidDet      & 0.752 & 0.236 & 0.873 & 0.153 & 0.901 & 0.152 & 0.904 & 0.174 & 0.935 & 0.154 & 0.930 & 0.083 \\
		DroidDetector & 0.526 & 0.384 & 0.826 & 0.625 & 0.937 & 0.813 & 0.865 & 0.757 & 0.751 & 0.467 & 0.763 & 0.506 \\
		HMMDetector   & 0.759 & 0.600 & 0.775 & 0.588 & 0.804 & 0.589 & 0.794 & 0.536 & 0.733 & 0.506 & 0.784 & 0.562 \\
		ICCDetector   & 0.612 & 0.180 & 0.732  & 0.120  & 0.780 & 0.096 & 0.797 & 0.156 & 0.832 & 0.141 & 0.833 & 0.055 \\
		MaMaDroid     & 0.756 & 0.199 & 0.879 & 0.136 & 0.923 & 0.159 & 0.923 & 0.187 & 0.901 & 0.145 & 0.895 & 0.076 \\ \hline
	\end{tabular}
\end{table*}

\begin{table*}[tb]
\centering
\caption{Decisions made by baseline detectors when dealing with greyware samples from the period 2012-2015, for different VTD values. G stands for the ratio of identifications as goodware, while M is the ratio of identifications as malware}\label{greyware}
	\begin{tabular}{c|cc|cc|cc|cc|cc|cc|cc}
		& \multicolumn{2}{c|}{\textbf{VTD=1}} & \multicolumn{2}{c|}{\textbf{VTD=2}} & \multicolumn{2}{c|}{\textbf{VTD=3}} & \multicolumn{2}{c|}{\textbf{VTD=4}} & \multicolumn{2}{c|}{\textbf{VTD=5}} & \multicolumn{2}{c|}{\textbf{VTD=6}} & \multicolumn{2}{c}{\textbf{1\textless=VTD\textless=6}} \\ \hline
		\textbf{Method} & \textbf{G} & \textbf{M} & \textbf{G} & \textbf{M} & \textbf{G} & \textbf{M} & \textbf{G} & \textbf{M} & \textbf{G} & \textbf{M} & \textbf{G} & \textbf{M} & \textbf{G} & \textbf{M} \\ \hline
		AndroDyalisis & 0.504 & 0.496 & 0.356 & 0.644 & 0.272 & 0.728 & 0.307 & 0.693 & 0.275 & 0.725 & 0.200   & 0.800   & 0.319                        & 0.681 \\
		BasicBlocks   & 0.675 & 0.325 & 0.452 & 0.548 & 0.394 & 0.606 & 0.333 & 0.667 & 0.248 & 0.752 & 0.248 & 0.752 & 0.392                        & 0.608 \\
		Drebin        & 0.696 & 0.304 & 0.460  & 0.540  & 0.379 & 0.621 & 0.323 & 0.677 & 0.251 & 0.749 & 0.148 & 0.852 & 0.376                        & 0.624 \\
		DroidDet      & 0.700   & 0.300   & 0.437 & 0.563 & 0.336 & 0.664 & 0.297 & 0.703 & 0.221 & 0.779 & 0.159 & 0.841 & 0.358                        & 0.642 \\
		DroidDetector & 0.629 & 0.371 & 0.691 & 0.309 & 0.607 & 0.393 & 0.659 & 0.341 & 0.607 & 0.393 & 0.536 & 0.464 & 0.622                        & 0.379 \\
		HMMDetector   & 0.415 & 0.585 & 0.296 & 0.704 & 0.311 & 0.689 & 0.259 & 0.741 & 0.255 & 0.745 & 0.163 & 0.837 & 0.283                        & 0.717 \\
		ICCDetector   & 0.683 & 0.317 & 0.460  & 0.540  & 0.386 & 0.614 & 0.381 & 0.619 & 0.295 & 0.705 & 0.240  & 0.760  & 0.408                        & 0.593 \\
		MaMaDroid     & 0.677 & 0.323 & 0.423 & 0.577 & 0.366 & 0.634 & 0.313 & 0.687 & 0.217 & 0.783 & 0.137 & 0.863 & 0.356                        & 0.645 \\ \hline
	\end{tabular}
\end{table*}

\begin{table}[tb]
\centering
\caption{Label swaps for goodware (G), greyware (X) and malware (M); between annotations of samples in the master dataset using information contained in AndroZoo (rows) and VirusTotal reanalysis reports (columns)}\label{swaps}
	\begin{tabular}{@{}cc|r|r|r}
		& & \multicolumn{3}{c}{\textbf{\theadb{VirusTotal\\reanalysis}}} \\ 
		& & \textbf{G} & \textbf{X} & \textbf{M} \\ \hline
		\multirow{3}*{\rotatebox{0}{\textbf{\theadb{AndroZoo\\analysis}}}} 
		& \textbf{G} & $8\,563$ &  $954$ &   $83$\\
		& \textbf{X} & $2\,613$ & $4\,284$ & $2\,703$\\
		& \textbf{M} &   $17$ &  $663$ & $8\,920$\\ \hline
	\end{tabular}
\end{table}

\subsection{Imbalance Scenario}

According to previous literature, around 10\% of apps are actually malware \cite{pendlebury2019tesseract}. Similarly, in this scenario we assume that there will be a strong imbalance between malware and goodware. In order to analyze the behavior of the different detectors in this context, the unbalanced dataset is used. It is worth noting that making the test set balanced (or unbalanced) does not harm the capabilities of the model, but only has an effect in some performance metrics. In this sense, the problem of imbalance in the test set is easily solved by using suitable metrics, such as the Kappa or A$_{mean}$, that do not conceal the errors for the minority class \cite{molinacoronado2020survey}.

Table \ref{tab:real_ratio} shows the results obtained for detectors under this scenario. The use of unbalanced data for training results in a reduction in the proportion of correctly classified malware (TPR), with respect to the results obtained using the baseline configuration (see Table~\ref{tab:baseline}). Imbalance in the training data is translated into less malware information provided to algorithms, making it difficult for detectors to learn the characteristics of this class of samples. On average the performance of most methods decreased about $12$\% as measured by their A$_{mean}$ values. Among the best scoring methods in the baseline scenario, MaMaDroid is one of the methods that suffered a more significant decrease in this scenario, with a reduction of $18$\% in its A$_{mean}$ value.

\begin{table}[bt]
	
	\centering
	
	\caption{Performance of detectors trained with 1:10 malware/goodware ratio}
	
	\resizebox{\linewidth}{!}{%
		\begin{tabular}{c|c|c|c|c|c|c}
			
			\textbf{Method} & \textbf{TPR} & \textbf{FPR} & \textbf{Precision} & \textbf{F1} & \textbf{A$_{mean}$} & \textbf{Kappa} \\ \hline
			
			AndroDialysis & 0.373 & 0.032 & 0.543 & 0.442 & 0.670 & 0.396 \\ 
			
			BasicBlocks & 0.460 & 0.017 & 0.734 & 0.565 & 0.721 & 0.531 \\ 
			
			
			Drebin & 0.740 & 0.017 & 0.816 & 0.776 & 0.861 & 0.754 \\ 
			
			DroidDet & 0.686 & 0.013 & 0.837 & {0.754} & {0.836} & {0.731} \\ 
			
			DroidDetector & 0.266 & 0.162 & 0.145 & 0.188 & 0.552 & 0.076 \\ 
			
			HMMDetector & 0.106 & 0.010 & 0.516 & 0.176 & 0.548 & 0.149 \\ 
			
			ICCDetector & 0.520 & 0.013 & 0.795 & 0.629 & 0.753 & 0.599 \\ 
			
			MaMaDroid & 0.513 & 0.008 & 0.865 & 0.644 & 0.752 & 0.617 \\ \hline 
			
		\end{tabular}
		
		}%
	
	\label{tab:real_ratio}
	
\end{table}

\subsection{Evasion Scenario}

This scenario is devoted to test the robustness of ML-based malware detectors under conditions where attackers attempt to bypass detection using obfuscation. Our aim with this section is to illustrate the need of carrying out the security analysis of detectors. There are many evasion techniques targetting the different phases of the malware detection process that have not been included in this study to keep the size and scope manageable, but deserve further analysis, e.g., sandbox detection \cite{vidas2014evading} or adversarial attacks against ML models \cite{papernot2017practical}. We focus our evaluation in obfuscation\footnote{Obfuscation involves code modifications to hinder the static analysis of apps without affecting their functionality. From the point of view of malware creators, obfuscation hampers the extraction of the features that may be indicative of malicious behaviors in order to fool detectors \cite{suarez2018eight}.} because it is a classical evasion technique that is commonly overlooked by the authors of proposals \cite{hammad2018large}. We selected a set of obfuscation strategies that are commonly used in the wild to hide Android malware behaviors \cite{dong2018understanding}. For each obfuscation strategy considered, as well as for the combination of all of them, a new obfuscated dataset is obtained by: (1) randomly sampling 10\% of the apps from 2012 to 2015 in our master dataset and, (2) applying the required transformations to such apps using the ObfusAPK \cite{aonzo2020obfuscapk} and the AAMO \cite{preda2017testing} tools. These transformations are: 

\begin{itemize}
	
	\item Renaming. The original name and identifiers of user-defined classes, fields and methods are changed by meaningless strings.
	
	\item Changes in the structure of the code. This form of obfuscation includes: (1) call indirections to add an intermediate function that calls the function originally present in the code, (2) insertion of \emph{goto} instructions, (3) inversion of conditionals to modify the execution flow of the app, (4) insertion of junk code, and (5) reflection\footnote{Reflection is a feature of some programming languages that allows an executing program to examine or ``introspect'' upon itself, and manipulate internal properties of the program.} to hide function calls to internal code and to the Android framework (APIs).
	
	\item Encryption. This technique involves: (1) the generation of a encryption/decryption random key, (2) the encryption of native libraries, strings and assets, and (3) the insertion of a decryption code which is called from every part of the app where these resources are requested.
	
\end{itemize}

The results are summarized in Table~\ref{tab:evasion}. On average, when considered separately, the most successful strategies for evading detection are changes in code structure and encryption. However, generally speaking, individual obfuscation strategies are not as effective as a combination of all of them. Most detectors are prone to misclassify obfuscated malware as goodware, as shown by the decrease in their TPR values for the combined scenario. In this regard, the least affected model is Drebin, being able to identify 79\% of the obfuscated malware, with 13\% false positives. Others, such as DroidDet and MaMaDroid obtained more moderate performance figures, both identifying around 70\% of the obfuscated samples according to their A$_{mean}$ values when using the three obfuscation techniques in combination.

\begin{table*}[tb]
	
	\centering
	
	\caption{Performance of detectors with obfuscated apps for the evasion scenario}
	
	\resizebox{\linewidth}{!}{%
		
		\begin{tabular}{c|c|c|c||c|c|c||c|c|c||c|c|c}
			& \multicolumn{3}{c||}{\textbf{\theadb{Renaming (Rn)}}} & \multicolumn{3}{c||}{\textbf{\theadb{Changes in code\\structure (Co)}}} & \multicolumn{3}{c||}{\textbf{\theadb{Encryption\\(Enc)}}}   & \multicolumn{3}{c}{\textbf{Rn+Co+Enc}}  \\\hline 
			\textbf{Method} & \textbf{TPR} & \textbf{FPR} & \textbf{A$_{mean}$} & \textbf{TPR} & \textbf{FPR} & \textbf{A$_{mean}$} & \textbf{TPR} & \textbf{FPR} & \textbf{A$_{mean}$} & \textbf{TPR} & \textbf{FPR} & \textbf{A$_{mean}$} \\\hline
			AndroDialysis & 0.793 & 0.288 & 0.752 & 0.376 & 0.072 & 0.652 & 0.431 & 0.074 & 0.678 & 0.459 & 0.086 & 0.686 \\
			BasicBlocks & 0.624 & 0.163 & 0.73 & 0.211 & 0.631 & 0.29 & 0.418 & 0.137 & 0.64 & 0.172 & 0.095 & 0.538\\
			Drebin & 0.95 & 0.012 & 0.968 & 0.968 & 0.025 & 0.971 & 0.738 & 0.138 & 0.799 & 0.796 & 0.136 & 0.829 \\ 
			DroidDet & 0.931 & 0.028 & 0.951 & 0.560 & 0.024 & 0.767 & 0.934 & 0.021 & 0.956 & 0.586 & 0.029 & 0.778 \\
			DroidDetector & 0.154 & 0.216 & 0.469 & 0.049 & 0.102 & 0.473 & 0.133 & 0.126 & 0.503 & 0.157 & 0.211 & 0.473 \\
			HMMDetector & 0.895 & 0.489 & 0.702 & 0.892 & 0.786 & 0.553 & 0.802 & 0.684 & 0.558 & 0.911 & 0.779 & 0.565 \\
			ICCDetector & 0.246 & 0.033 & 0.606 & 0.667 & 0.023 & 0.822 & 0.668 & 0.023 & 0.822 & 0.279 & 0.024 & 0.627 \\
			MaMaDroid & 0.733 & 0.034 & 0.848 & 0.737 & 0.042 & 0.845 & 0.92 & 0.037 & 0.941 & 0.533 & 0.048 & 0.743 \\ \hline
	\end{tabular}
	
}%

\label{tab:evasion}

\end{table*}

\subsection{Evolution Scenario}

Both malware and goodware evolve over time, i.e., do not follow a stationary distribution. Therefore, it is logical to think that static detectors trained with apps for a certain period of time will not necessarily work well with more recent apps. To prove this assumption, the models trained using the balanced configuration (with data from the period 2012-2015) are tested with goodware and malware obtained between 2016 and 2019.

\begin{figure}[h]
\centering
\begin{subfigure}[hb]{\linewidth}
	\centering
	\includegraphics[width=\linewidth]{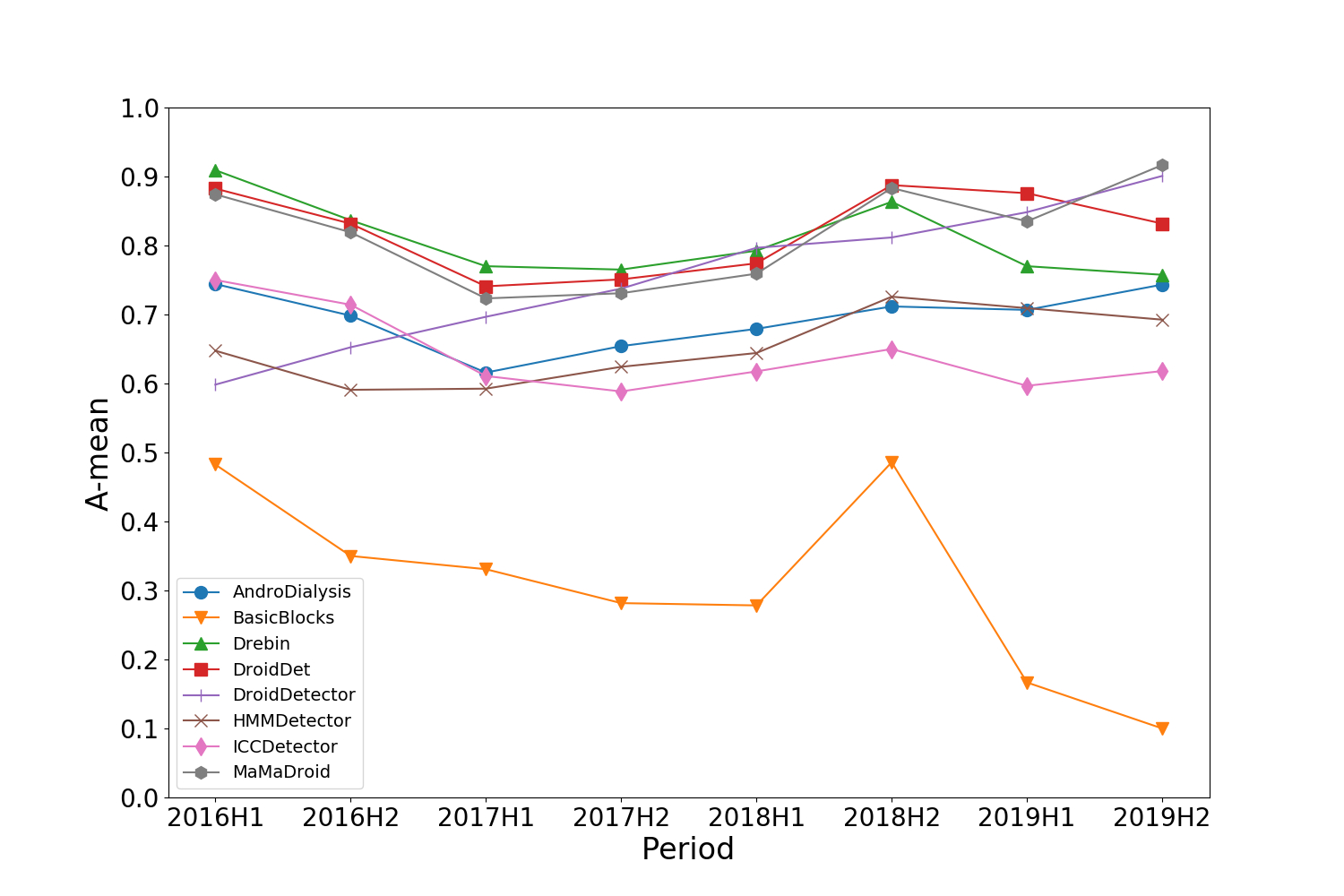}
	\caption{A$_{mean}$ values}
	\label{fig:amean}
\end{subfigure}
\begin{subfigure}[hb]{\linewidth}
	\centering
	\includegraphics[width=\linewidth]{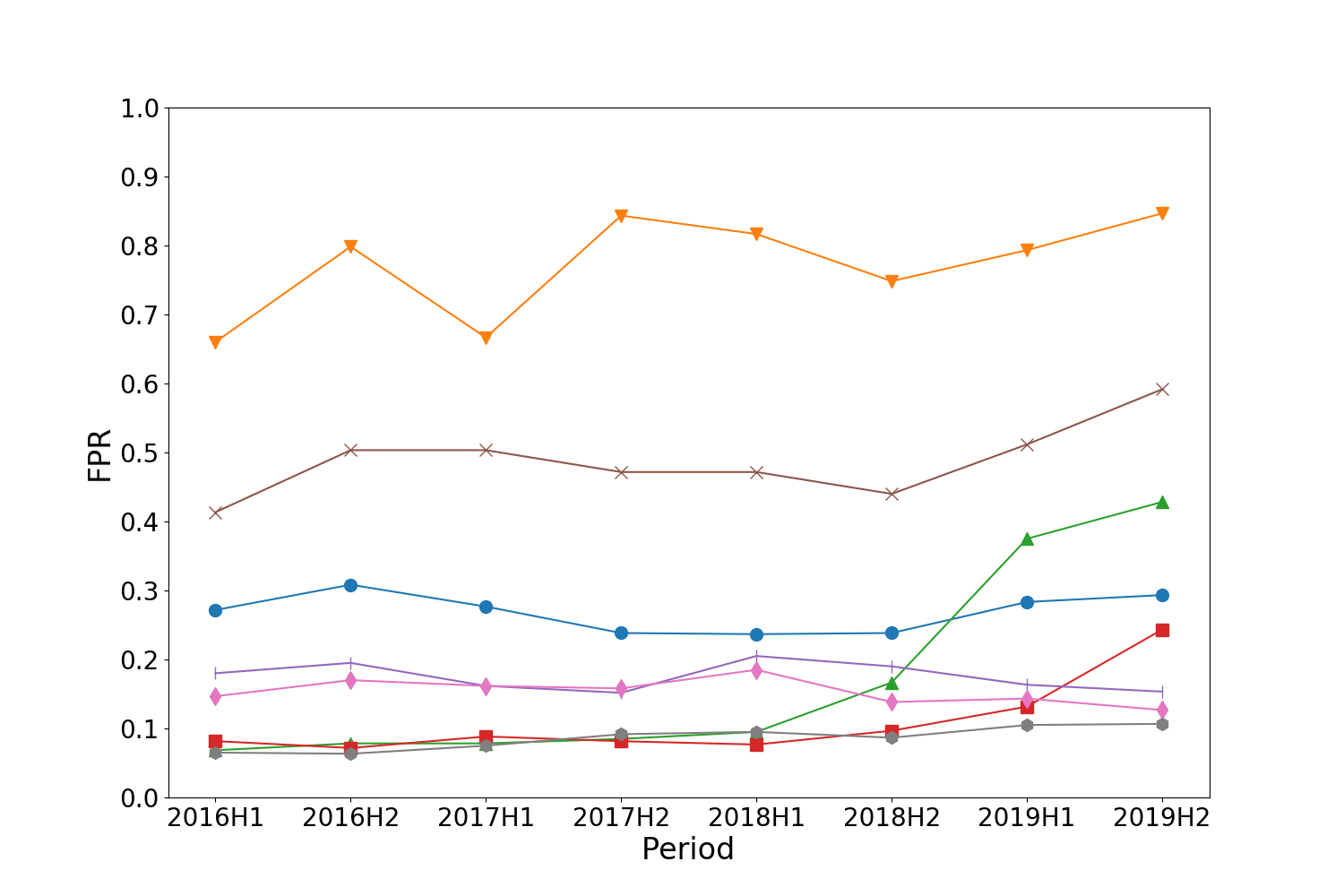}
	\caption{FPR values}
	\label{fig:fpr}
\end{subfigure}
\begin{subfigure}[hb]{\linewidth}
	\centering
	\includegraphics[width=\linewidth]{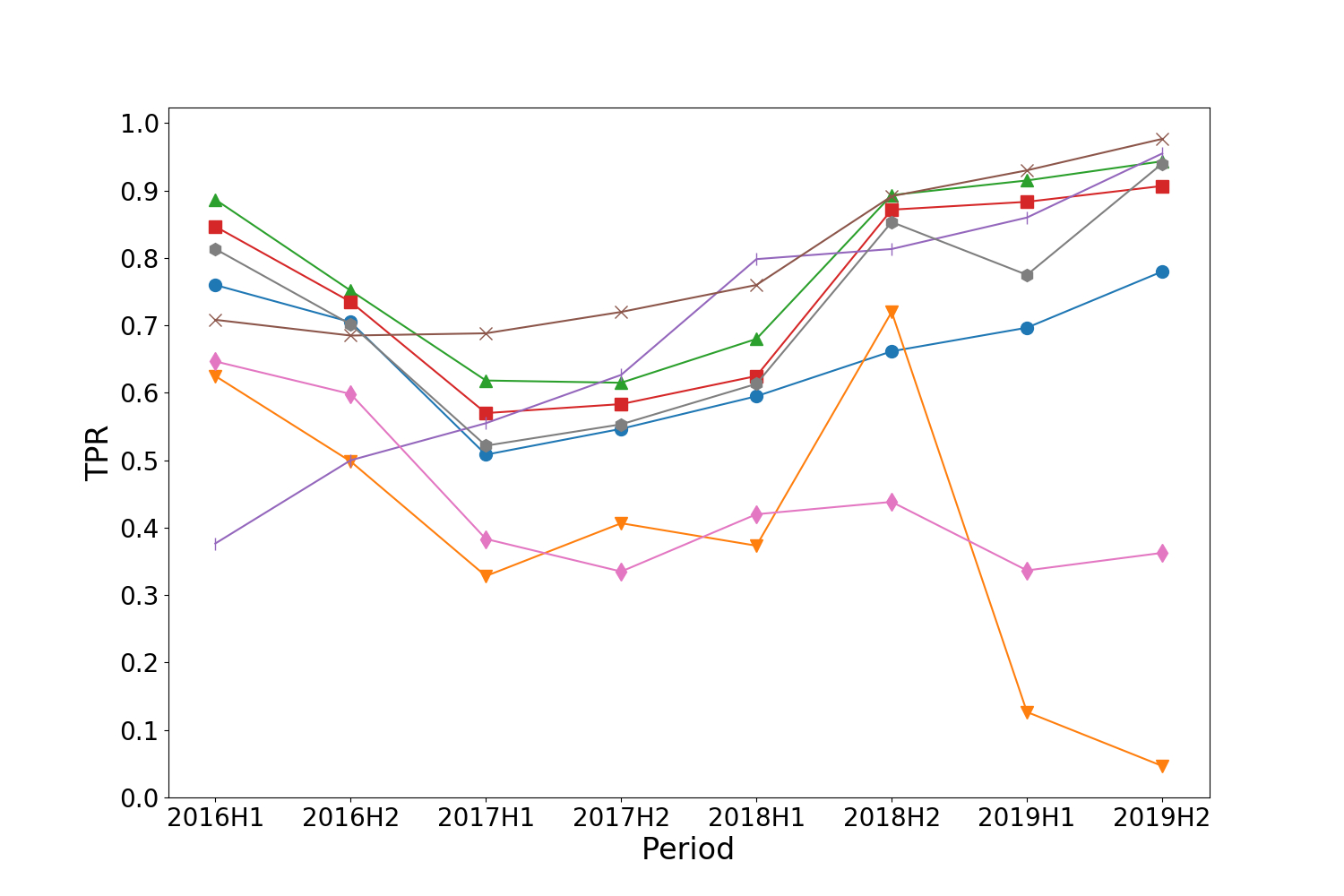}
	\caption{TPR values}
	\label{fig:tpr}
\end{subfigure}
\caption{Evolution of the performance of baseline detectors for the period 2016-2019}\label{fig:evolution}
\end{figure}

The overall performance of most of the methods under this scenario is characterized by an increment of both false positives and false negatives (see Table~\ref{tab:dynamic}). Figure~\ref{fig:amean} provides a more detailed view of the evolution of the TPR, FPR and A$_{mean}$ values throughout the period between 2016 and 2019. As can be seen, the performance of detectors changes in a notable manner from one period to another. Our hypothesis is that the popularity of some malware families decreases with time, and, at some point, they are replaced with newer families for which a detector may not have been trained. Also, at some point, an old behavior may become popular again, resulting in a sudden increase in the performance of that detector. This is why the lines in Figure~\ref{fig:tpr} do not follow clear, decreasing trends. In addition, contrary to what is stated in \cite{pendlebury2019tesseract}, the incremental trend in the number of false alarms of detectors (see~Figure~\ref{fig:fpr}) indicates that goodware also changes its behavior over time. At any rate, our observations demonstrate the non-stationary nature of malware and goodware, and show how classic, batch-trained ML algorithms are not an appropriate solution for Android malware detection given the instability of their decisions.

\begin{table}[t]

\centering

\caption{Performance of baseline detectors using evaluation data between 2016 and 2019}

\resizebox{\linewidth}{!}{%
	\begin{tabular}{c|c|c|c|c|c|c}
		
		\textbf{Method} & \textbf{TPR} & \textbf{FPR} & \textbf{Precision} & \textbf{F1} & \textbf{A$_{mean}$} & \textbf{Kappa} \\ \hline
		
		AndroDialysis & 0.656 & 0.268 & 0.709 & 0.682 & 0.694 & 0.388 \\ 
		
		BasicBlocks & 0.390 & 0.771 & 0.336 & 0.361 & 0.309 & -0.381 \\ 
		
		
		Drebin & 0.787 & 0.171 & 0.820 & 0.804 & 0.808 & 0.616 \\ 
		
		DroidDet & 0.752 & 0.108 & 0.873 & 0.808 & 0.821 & 0.643 \\ 
		
		DroidDetector & 0.685 & 0.175 & 0.796 & 0.736 & 0.755 & {0.510} \\ 
		
		HMMDetector & 0.795 & 0.488 & 0.619 & 0.696 & 0.653 & 0.306\\ 
		
		ICCDetector & 0.440 & 0.153 & 0.741 & 0.552 & 0.643 & 0.286 \\ 
		
		MaMaDroid & 0.721 & 0.086 & 0.893 & {0.798} & {0.817} & {0.635} \\ \hline 
		
	\end{tabular}
	
	}%

\label{tab:dynamic}

\end{table}

\section{Discussion: Towards a Realistic Framework for Malware Detection}\label{realistic}

Based on the factors analyzed in the previous sections, this section presents a collection of ideas or recommendations to consider when designing and evaluating a realistic proposal for Android malware detection. 

Undoubtedly, one of the main problems encountered when researching malware detectors for Android is the lack of reproducibility of many proposals in the literature. The datasets and code used in the experimental processes are rarely public, and the details provided in the papers are often not enough for a correct re-implementation of the methods. Therefore, the most important recommendation is that authors of future works ensure that their work and experimental processes are fully reproducible. In addition to this general aspect, the experiments conducted in the previous section have highlighted the importance of using adequate datasets when training models and assessing their performance. Indeed, as we have seen, depending on the datasets used, the same model can show near-perfect performances or be almost irrelevant. Thus, we can conclude that the datasets and experimental scenarios considered in the literature are unrealistic and should be revised. 

Related to the previous statements, our analysis has shown that the presence of highly similar apps in the datasets influences the performance of models. This also opens the possibility to perform evasion attacks against classifiers, for example, by developing malware apps with snippets of code from minor malware variants. During training, models tend to focus on large groups of duplicates, instead of trying to generalize the variety of apps in the data. During model assessment, duplicates are the cause of misleading performance indicators that tend to overestimate the detection ability of models. We can conclude that training and evaluation efforts should leverage on datasets without duplicates to improve and demonstrate the generalization capacity of models \cite{hastie2009elements}. Training should be carried out after balancing the representatives within each of the classes, i.e., between groups of similar samples; to avoid biases and exploit the generalization capacity of ML models. Also, contextual evaluations that take into account the prevalence of different malware families or groups are desirable to describe the actual reasons behind the performance of detectors. The removal of duplicates has an additional advantage, as the implicit reduction in the dimensionality of the data is useful to speed up the labeling, training and evaluation processes.

In relation to the labeling of the apps, we have seen that VirusTotal is widely used, but relying only on the VTD reported by this tool entails some risks \cite{zhu2020measuring}. To begin with, there is a lack of consensus on the labeling of the apps and on the inclusion of greyware in the training and evaluation of models. We should bear in mind that the line between malware and goodware is not clearly defined. As we demonstrated, the choice of the labeling criteria highly influences the generated model and the detection performance, simplifying the problem as the VTD is increased. This lack of consensus also results in the omission of a large amount of greyware which lies in between. However, such apps are present in the Android app ecosystem \cite{AndroidSec,arp2020and}, so, realistic proposals should not ignore these type of apps but include them in their training and evaluation processes. The nature of such apps may be uncertain. In this regard, additional information such as the details included in the VirusTotal reports can also be used to support more sophisticated labeling techniques, identifying particular functionalities like advertising, bundling, etc. which could provide valuable information to the user about a given (greyware) app. This would help to create a taxonomy of the type of apps present in the Android ecosystem, and to reach consensus on what is goodware or malware. It would also help to determine the usefulness of a third category of ``potentially unwanted apps'' \cite{nortonpuas} and its consideration as a third class in the data. In addition, to avoid the use of fixed thresholds over the VTD for labeling, and as an alternative to binary or three-class classifiers, new detection proposals could explore regression models that provide a risk/maliciousness metric, multi-step learning approaches \cite{daoudi2022two}, unsupervised detectors that return a degree of dissimilarity with respect to the benign class \cite{mahindru2021semidroid}, or semi-supervised methods that do not require fully labeled datasets \cite{van2020survey}.

Also regarding labeling, a more general problem is that not all antivirus engines in VirusTotal are equally reliable, with some of them being correlated or specialized in specific types of malware \cite{miller2016reviewer}. In addition, two engines from the same vendor but specific for different platforms may differ \cite{salem2021maat}. Although reliable labels can be derived from the VTD \cite{zhu2020measuring}, setting simple thresholds to the VTD assumes that all antivirus engines are equally reliable in all situations and makes no distinctions between them. In order to overcome these aspects, techniques such as crowd learning \cite{whitehill2009whose}, which measure the relevance of the different antivirus engines present in VirusTotal, could be used. Differences among vendors are even more noticeable when obtaining family labels from malware, since anti-virus firms do not share a common nomenclature. In this regard, despite that some useful tools are available for agreeing on family labels from VirusTotal reports \cite{sebastian2020avclass2}, we believe that there is still a lot of work to be done. For example, proposing a standard taxonomy of malicious behaviors that would allow a common specification of malware.

As a last problem with labeling, we have seen in the experimentation that the VTD, and thus, the labeling, changes over time \cite{zhu2020measuring}. Weakly labeled data and changes in the labeling can hinder the generation of robust classifiers, leading to detection errors. As a solution, the use of the VTD for labeling, on its own, should be abandoned and replaced by more sophisticated methods, for example, that incorporate informative features provided in the analysis reports of VirusTotal \cite{salem2021maat}. Active learning may also be a useful tool in this sense. This technique allows the best representatives of each class to be selected, reducing the effort of manual labeling \cite{huang2010active}. Nonetheless, using sophisticated methods for labeling might become cumbersome and costly, and of course, their reliability should be validated. 

Another important aspect is that malware and goodware apps are found in different proportions in the wild, depending on the source of the apps, but malware being the minority class \cite{pendlebury2019tesseract}. Ideally, a detector should perform well when analyzing apps regardless of the source of the samples. Thus, proposals should be trained and tested assuming unbalanced datasets. Additionally, the adoption of suitable performance metrics for unbalanced scenarios, such as the kappa and A-mean metrics, should be contemplated as opposed to using others, such as the accuracy, that do not reflect the real performance of detectors in these contexts. We have seen that classical ML supervised classification algorithms used for malware detection do not properly manage imbalance, as they expect data to be equally balanced in order to build a robust model \cite{he2009learning}. As such, the use and design of specific ML classifiers for unbalanced scenarios is an open research line, which could be promising in this area of study. Some examples include cost-based classification methods, and subsampling and oversampling methods \cite{fernandez2018learning}. In addition, depending on the particular scenario in which the model will be applied, and on the interest of the practitioner (for example, reducing the number of FPs), different loss functions could be used in the training phase. As alternative approaches, anomaly detection algorithms may also be considered \cite{molinacoronado2020survey}. 

Next, detectors have to work in a hostile scenario where attackers will try to trick them to infect a system. These are one of the first lines of defence, so security of malware detectors is important. Despite that our evaluation was limited to simple evasion attacks based on the use of obfuscation techniques, we evidenced that most detectors are not designed with security in mind and are, in fact, vulnerable to attacks. Appropriate assessments of the robustness of detectors should consider all the steps involved in the malware detection process. Novel evaluation methodologies that take into account the latest technology and attacks are needed and represent a promising future line of research. These should consider the evaluation of approaches to attacks targeting the analysis phase of apps, such as sandbox detection \cite{yokoyama2016sandprint}; or, to attacks that target ML algorithms, such as adversarial learning \cite{zhang2019adversarial}. In this regard, the proposal of robust ML algorithms against novel attacks is a very interesting research line \cite{demontis2017yes}. However, we should not forget that cutting-edge detectors will also need to counteract classical evasion attacks. Thus, the selection and design of feature sets that are most useful to detect obfuscated malware behaviors \cite{garcia2018lightweight}, as well as, the improvement of app analysis tools, are still promising research areas. Therefore, obfuscated samples are also essential in testing datasets. As for evasion techniques and tools, in this paper we have considered only a few; however, a broader set of techniques and tools that implement them need to be thoroughly analyzed in future work.

Finally, the malware detection problem is non-stationary, i.e., classes evolve over time and rarely show constant characteristics. A realistic detector should be able to cope with the non-stationary nature of Android apps. We demonstrated that the analyzed detectors, which are based on classical ML classifiers, are not able to manage such changes. A simple solution to this limitation could be to apply periodic retraining procedures to obtain up-to-date detectors \cite{molinacoronado2020survey}. However, depending on the complexity of the algorithms, this process could increment the cost of building and maintaining detectors. Also, the cost of obtaining a sufficient number of labeled samples at a time should not be obviated. Some authors have focused on slowing-down the aging of models and reduce update efforts by finding similar API functions \cite{zhang2020enhancing}. As a more promising ML alternative for dynamic scenarios, online or stream learning algorithms have been proposed \cite{narayanan2017context}, some of which weight the instances based on their prevalence or age. These algorithms can automatically manage the changes in the distribution of the data so that they are able to adapt the models without intervention whenever these changes occur \cite{lu2018learning}. Accordingly, the suitability of these methods for malware detection needs to be explored further. Another option would be to consider anomaly detectors or hybrid misuse-anomaly detectors \cite{molinacoronado2020survey} as they may be more capable of dealing with the evolution in the data and zero-day attacks.

\section{Conclusions}\label{sec:concluding}

One of the main conclusions of this work is that authors of ML-based Android malware detectors tend to be very optimistic when designing and evaluating their systems, ignoring factors such as the presence of duplicates in the datasets, the lack of robust labeling methods, the presence of greyware, the imbalance between malware and goodware, the existence of apps trying to evade detection and the evolution of apps. Our evaluation work has shown how these factors substantially affect the performance that can be achieved with the tested detectors. We have seen, therefore, that malware detectors are not ready for deployment in real environments. Another important problem we have studied is the lack of a common design and evaluation framework, stemming, among other things, from the unavailability of codes and standardized and appropriate datasets. This fact greatly complicates the reproducibility of experiments. Our contribution in this regard has been to release our data and codes so that they can be used by other researchers. Finally, we have included in this paper a number of ideas that can contribute to the design of malware detectors which are able to operate in realistic deployment scenarios.

\section*{CRediT authorship contribution statement}
\textbf{Borja Molina-Coronado:} Conceptualization, Methodology, Software, Formal analysis, Investigation, Writing - Original Draft, Writing - Review \& Editing \textbf{Usue Mori:} Conceptualization, Methodology, Writing - Review \& Editing \textbf{Alexander Mendiburu:} Conceptualization, Methodology, Writing - Review \& Editing \textbf{Jose Miguel-Alonso:} Conceptualization, Methodology, Writing - Review \& Editing

\section*{Acknowledgments}

This work has received support from the following programs: PID2019-104966GB-I00AEI (Spanish Ministry of Science and Innovation), IT-1504-22 (Basque Government), KK-2021/00095 and KK-2021/00065 (Elkartek projects SIGZE and ALUSMART supported by the Basque Government). Borja Molina-Coronado holds a predoctoral grant (ref.~PRE\_2020\_2\_0167) by the Basque Government.

	\small
	
	\bibliography{references}
	
	\bibliographystyle{unsrtnat}

\end{document}